\documentclass[article]{elsarticle-anonymous}

\usepackage{natbib}
\usepackage{latexsym,amssymb,amsmath,ae,aeguill}
\usepackage[french,english]{babel}
\usepackage{graphics,graphicx,color}
\usepackage{subfigure}
\usepackage{epsfig}
\usepackage{epstopdf}
\usepackage{bm}
\usepackage{multiequations}
\usepackage{url}

\newcommand*{\be}{\begin{equation}}
\newcommand*{\ee}{\end{equation}}
\newcommand*{\bea}{\begin{eqnarray}}
\newcommand*{\eea}{\end{array}}
\newcommand*{\bal}{\begin{align}}
\newcommand*{\eal}{\end{align}}
\newcommand*{\bme}{\begin{multiequations}}
\newcommand*{\eme}{\end{multiequations}}

\newcommand\Rey{\mbox{\textit{Re}}}  
\newcommand{\reel}{\mathbb R}
\newcommand{\R}{\mathbb{R}}

\newcommand{\N}{\mathbb{N}}
\newcommand{\defin}{\stackrel{\scriptscriptstyle\triangle}{=}}
\newcommand{\inlaw}{\stackrel{\scriptscriptstyle{\cal L}}{=}}

\newcommand{\B}{\boldsymbol{B}}
\newcommand{\bsigma}{\boldsymbol{\sigma}}
\newcommand{\XX}{\boldsymbol{X}}
\newcommand{\xx}{\boldsymbol{x}}
\newcommand{\zz}{\boldsymbol{z}}
\newcommand{\YY}{\boldsymbol{Y}}
\newcommand{\nab}{\boldsymbol{\nabla}}
\newcommand{\car}{1\!\mathsf{I}}

\providecommand\bcdot{\boldsymbol{\cdot}}
\newcommand{\transp}{^{\scriptscriptstyle T}}

\newcommand{\Exp}{\mathbb{E}}
\newcommand{\Id}{\mathbb{I}}
\newcommand{\mbf}[1]{\ensuremath{\mathbf{#1}}}
\newcommand{\mbs}[1]{\ensuremath{\boldsymbol{#1}}}

\def\a{\mbox{$\boldsymbol{a}$}}
\def\x{\mbox{$\boldsymbol{x}$}}
\def\f{\mbox{$\boldsymbol{f}$}}

\def\S{\mbox{$\boldsymbol{S}$}}

\newcommand{\w}{\boldsymbol{w}}
\newcommand{\we}{\widetilde{\;\w\;}}
\newcommand{\yy}{\boldsymbol{y}}

\newcommand{\dif}{{\mathrm{d}}}

\newcommand\eg{e.g.\ }

\begin{document}


\title{Stochastic representation of the Reynolds transport theorem: revisiting large-scale modeling}


\author[skh]{S. Kadri Harouna}
\author[em]{E. M\'emin\corref{cor1}}
\ead{etienne.memin@inria.fr}
\address[skh]{Laboratoire Math\'ematiques, Image et Applications (MIA), Universit\'e de La Rochelle, Avenue Michel Cr\'epeau 17042 La Rochelle, France}
\address[em]{INRIA Rennes, Campus Universitaire de Beaulieu, 35042 Rennes cedex, France}

\date{\today}

\begin{abstract}
We explore the potential of a formulation  of the  Navier-Stokes equations  incorporating a random description of the small-scale velocity component.  This model, established  from a version of the Reynolds transport theorem adapted to a stochastic representation of the flow, gives rise to a large-scale description of the flow dynamics in which emerges an anisotropic subgrid tensor, reminiscent to the Reynolds stress tensor, together with a drift correction due to an inhomogeneous turbulence. The corresponding subgrid model, which depends on the small scales velocity variance, generalizes the Boussinesq eddy viscosity assumption. However, it is not anymore obtained from an analogy with molecular dissipation but ensues rigorously from the random modeling of the flow. This principle allows us to propose several  subgrid models defined directly on the resolved flow component.  We assess and compare numerically those models  on a standard Green-Taylor vortex flow at Reynolds 1600. The numerical simulations, carried out with an accurate  divergence-free scheme, outperform classical large-eddies formulations and provides a simple demonstration of the pertinence of the proposed large-scale modeling.  
\end{abstract}


\maketitle

\section{Introduction}
%
The large-scale modeling of fluid flow dynamics  remains nowadays a major research issue  in fluid mechanics or in geophysics despite an enormous research effort since the first investigations on the subject 150 years ago \citep{Boussinesq77}. The research themes behind this topic cover fundamental issues such as turbulence modeling and the analysis of fully developed turbulent flows, but also more applicative research problems related to the definition of practical numerical methods for the simulation of complex flows. In this latter case the difficulty consists in setting up a reliable modeling of the large-scale dynamics in which the contribution of unresolved small-scale  processes are explicitly taken into account. For the Navier-Stokes equations, the problem is all the more difficult that the spatial and temporal scales are tightly interacting together. 

The neglected processes include, among others things, the action of the unresolved motion scales, complex partially-known forcing, an incomplete knowledge of the boundary conditions and  eventual numerical artifacts. Such unresolved processes must be  properly taken into account to describe accurately the energy transfers and to construct  stable numerical simulations. In real world  situations, the complexity  of the involved phenomenon  prevents the use of  an accurate -- but inescapably expensive -- deterministic modeling. We advocate instead the use of a stochastic modeling. 

 Within this prospect, we aim at describing the missing contributions through random fields encoding a flow component only in a probability distribution sense. Those variables correspond to the discrepancies or errors between the dynamical model and the actual dynamics. Their modeling is of the utmost importance  in geophysics, either for data assimilation or forecasting issues. In both cases, an accurate  modeling of the flow errors dynamics enables to maintain an ensemble of flow configurations with a sufficient but also meaningful diversity. 

Small-scale processes  are responsible both for an energy dissipation but also for local backscattering of energy \citep{Piomelli91}. The introduction  of random variables in the flow dynamics has been considered by several authors, as it constitutes an appealing mechanism for the phenomenological modeling of intermittent processes associated to the inverse energy cascade \citep{Leith90,Mason92,Schumann95}. Recently those models have regained a great interest for the modeling of geophysical flows \citep{Buizza99,Grooms13,Majda99,Majda03,Shutts05} in climate sciences (see also the thematic issue \citep{Palmer08} or the review \citep{Franzke15}). 

Numerous turbulence models proposed in the context of Large Eddies  Simulations (LES) and Reynolds Average Simulations  (RANS) introduce {\em de facto} an eddy viscosity assumption to model the energy dissipation due to unresolved scales \citep{Deardorff70,Lilly66,Meneveau00, Smagorinsky63,Sagaut05}. This concept dates back to the work of Boussinesq \citep{Boussinesq77} and Prandtl  \citep{Prandtl25}. It relies on the hypothesis that the energy transfer from the resolved scales to the subgrid scales can be described in a similar way to the molecular viscosity mechanism. It is therefore not at all related to any uncertainty or error quantities. In models dealing explicitly with a statistical modeling of the turbulent fluctuations there is thus some incoherency in representing directly the dissipative mechanism attached to random terms through an eddy viscosity assumption.  In this work we will not make use of such an hypothesis. Instead, we will rely on a general diffusion expression  that emerges naturally from our formalism. 

 This subgrid model is properly derived from a general Lagrangian stochastic model of the fluid motion in which the fluid parcels displacement is decomposed in two components: a smooth differentiable (possibly random)  function  and a random field, uncorrelated in time but correlated in space.  Such a decomposition consists  in separating or  "filtering" a  rough velocity in a smooth slow time-scale  component and a fast oscillating velocity field representing the unresolved flow. Though there is, in general, no sharp time-scale separation in turbulent flows, the resolved velocity can be interpreted as a temporally coarse-grained component whereas the time-uncorrelated component stands for the small time-scale unresolved velocity. As a temporal smoothing  imposes implicitly a spacial smoothing, this separation can  be thus interpreted in terms of a LES filtering technique. Yet, the corresponding Eulerian formulation does not  ensue from a filtering procedure. It is thus not prone to errors associated to the violation of the commutation assumption between the filter and the spatial derivatives \citep{Ghosal95,Ghosal96}. Besides, those equations introduce an effective  advection related to the small-scale velocity inhomogeneity. This modified advection, empirically introduced  in Langevin models of particle dispersion \citep{Macinnes92}, corresponds exactly to a phenomenon, termed {\em turbophoresis}, related to the  migration of inertial particles  in regions of lower turbulent diffusivity \citep{Sehmel70}.

The large-scale representation of the Navier-Stokes equations on which we rely in this study are built from a stochastic version of the Reynolds transport theorem \citep{Memin14}. This modified Reynolds transport theorem, which constitutes here the  cornerstone of our large-scale dynamics representation, is presented in the following section. General invariance properties of  the corresponding large-scale dynamics such as scale and Galilean invariances are detailed in a comprehensive apppendix. 
In section \ref{Experiments} several novel subgrid tensors will be devised and compared on a standard Green-Taylor vortex flow \citep{Taylor-Green}. We will show that all the proposed schemes outperform the usual dynamic Smagorinsky subgrid formulation \citep{Germano91,Germano92,Lilly92,Smagorinsky63}. 
\section{Stochastic modeling of fluid flow dynamics}\label{NSU}
 Numerous methodological choices can be envisaged to devise stochastic representations of the Navier-Stokes equations. The simplest method considers additional random forcing to the dynamics. This is the choice that has been the most often performed since the work of Benssoussan \citep{Bensoussan-Temam-73}. Another choice, in the wake of Kraichnan's work \citep{Kraichnan59}, consists in closing the large-scale flow representation in the Fourier space by  relying on a Langevin equation \citep{Kraichnan70, Laval06, Leith71}. Obviously the frontiers between those two methodologies are sometimes fuzzy, and numerous works rely on both  strategies in order to devise the shape that should take the random variables  evolution law \citep{Laval06, Schumann95}. Lagrangian stochastic models based on Langevin equations have been also intensively used for turbulent dispersion \citep{Sawford86} or in probability density function (PDF) modeling of turbulent flows \citep{Haworth86,Pope00}. Those Lagrangian models, which require to model the drift and diffusion functions, lead to very attractive particle based representations of complex flows \citep{Pope94,Minier14}. They are nevertheless not adapted to global large-scale Eulerian representations of the flow dynamics.  
 
 In this work, we will rely on a different framework in specifying  the stochastic nature of the velocity from the very beginning as proposed in \citep{Memin14}. The basic idea is built on the assumption that the Lagrangian fluid particles displacement results from a smooth velocity component and a highly oscillating stochastic velocity component uncorrelated in time,
 \begin{equation}\label{sto-velocity}
\XX_t  = \XX_{t_0} + \int_{t_0}^{t}\w(\XX_s,s) \dif s +  \int_{t_0}^{t}\bsigma(\XX_s,s) \dif \B_s,
\end{equation}
 with the velocity components:
 \begin{equation}\label{sto-velocity}
\mbs U(\XX_t,t)  = \w(\XX_t,t) +  \dot{\mbs W}(\XX_t,t) .
\end{equation}
In this decomposition the first right-hand term is a smooth function of time associated to the  large-scale velocity component. The second term stands for the small-scale velocity field. It is a white noise velocity component defined from the (formal) time-derivative of the random field: $\dot{\mbs W}(\XX_t,t) =\bsigma(\XX_t,t) \frac{\dif}{\dif t}\B_t$. This random field is a three-dimensional centered Wiener process; it is thus uncorrelated in time but  can be anisotropic and inhomogeneous in space. Since  we focus in this study only on incompressible flows, the small-scale component is defined as a divergence-free random field; it is hence associated to a divergence-free diffusion tensor:
\begin{equation}
\label{incompressiblity cond sigma}
\nab\bcdot \bsigma=0.
\end{equation}
Analogously to the standard deterministic case, the derivation procedure from the physical conservation  laws of the Navier-Stokes equations is based primarily on the Reynolds transport theorem (RTT). 
\subsection{Stochastic Reynolds transport theorem}

The RTT provides the expression of the rate of change of a scalar function, $q$, within a material volume, ${\cal V}(t)$. For a stochastic flow (\ref{sto-velocity}) with an incompressible small-scale velocity component ($\nab\bcdot \mbs \sigma =0$), this expression derived in \citep{Memin14} (see also appendix \ref{appendix:srt} where a comprehensive derivation is provided for readers convenience), is given by:
\begin{multline}
\dif\int_{{\cal V}(t)} \!\!\!\!\!\!q \dif \x = \int_{{\cal V}(t)} \Biggl(\dif_t q +\Bigl[ {\nab} \bcdot \bigl(q (\underbrace{\w -\frac{1}{2} \nab\bcdot \mbs a}_{\widetilde{\;\w\;}})\bigr)\\
-\frac{1}{2} \sum_{i,j=1}^d \partial_{x_i}  (a_{ij} \partial_{x_j}q)\Bigr] \dif t  +\nab q \bcdot \bsigma \dif\B_t\Biggr)\dif \x. 
\label{Sto-R}~
\end{multline}
This modified RTT involves the time increment of the random scalar quantity $q$ (the differential of $q$ at a fixed point) instead of the time derivative. A diffusion operator emerges also naturally. For clarity's sake, this term is designated as ``subgrid stress tensor'' following the protocols of large eddies simulation (LES). However, its construction is quite different. It is not based on Boussinesq's eddy viscosity assumption nor on any structural turbulence models \citep{Sagaut05} but arises directly from stochastic calculus rules. It expresses the mixing process exerted on the scalar quantity by the fast oscillating velocity component. This diffusion term  is directly related to the small-scale component through the {\em variance tensor}, $\mbs a$, defined from the diagonal of the small-scale velocity covariance: 
\begin{equation*}
 \mbs a(\xx,t)\delta(t-t')\dif t= \Exp \left ( \left (\mbs \sigma(\xx,t) \dif\B_t \right ) \left ( \mbs \sigma(\xx,t') \dif\B_{t'} \right ) \transp \right ),
\end{equation*}
it can be checked that the variance tensor corresponds to an eddy viscosity term (with units in $m^2s^{-1})$.
This term plays thus a role similar to the eddy viscosity models introduced in classical large scale representations \citep{Bardina80,Gent90,Lilly92,Smagorinsky63} or to the spectral vanishing  viscosity \citep{Karamanos00,Pasquetti06,Tadmor89}. It is also akin to numerical regularization models considered in implicit models \citep{Aspden08,Boris92,Dairay16,Lamballais11}. Our approach is nevertheless more general as it does not rely on {\em a priori} fixed shape of the Reynold stress (\eg Boussinesq assumption) nor does it  presuppose a given numerical discrete scheme (\eg implicit models).

A corrective advection term, $\widetilde{\w}= \w -1/2 \nab\bcdot \mbs a$, appears also in the stochastic RTT formulation. This correction expresses the influence of the small scales inhomogeneity on the large ones. A drift is engendered from the regions associated with maximal variance (maximal turbulent kinetic energy - TKE) toward the area of minimum variance (e.g. minimal TKE). This corresponds to the {\em turbophoresis} phenomenon associated with turbulence inhomogeneity, which drives inertial particles toward the regions of lower diffusivity \citep{Brooke92,Caporaloni75,Sehmel70}. For homogeneous noise, the variance tensor is constant and this corrective advection does not come into play. 
It can be noted that this advection correction is of the same form as the one proposed in \citep{Macinnes92}.

Through this modified RTT, stochastic versions of the mass and momentum conservation equations can be (almost) directly derived. Incompressibility conditions can for instance be immediately deduced from the RTT applied to $q=1$ and the flow jacobian ($J$): 
\begin{multline}
\label{th_transport-formal}
\hspace{-0.5cm}
\int_{{\cal V} (t_0)}\!\!\!\!\!\!\!\!\dif (J(\XX_t(\xx),t)) \dif \xx
=
\dif \!\!\int_{{\cal V} (t)} \!\!\!\!\!\!\! \dif \xx 
= \\
\int_{{\cal V} (t)} \!\!\!\!\!\!\!\!\nab \bcdot \widetilde{\w} (\xx,t)\dif t \dif \xx 
=
\int_{{\cal V} (t_0)} 
\!\!\!\!\!\!\!\! \left(J \nab\bcdot \widetilde{\w}\right)  (\XX_t(\xx),t)
\dif t \dif \xx .
\end{multline}
Together with the incompressibility of the random term, the incompressibility condition reads thus:
\begin{equation}
\label{eq_incomp_sto}
\nab\bcdot \mbs \sigma = 0
\text{ and }
\nab\bcdot \widetilde{\w}
= 0.
\end{equation} 
In the case of an incompressible large-scale flow component, $\w$,  this reduces to:
\begin{equation}
\label{incompressibility}
\nab\bcdot \mbs \sigma= 0
\text{ and }
\nab\bcdot \w = \nab\bcdot \left( \nab\bcdot \mbs a \right) \transp=0.
\end{equation}
Note that for a divergence-free isotropic random field such as the Kraichnan model \citep{Kraichnan68}  the last condition is naturally satisfied, since this unresolved velocity component is associated with a constant variance tensor.
\subsection{Mass conservation}
 Applying the RTT to the fluid density, $\rho$, and canceling this expression for arbitrary volumes, we get the following  mass conservation constraint:
\begin{align}
\label{mass-conservation-1}
&\dif_t \rho_t + {\nab} \bcdot(\rho \we ) \dif t + \nab \rho\bcdot \bsigma \dif  \B_t = \frac{1}{2} \nab\bcdot   (\mbs a  \nab q)\dif t,\\
&\we = \w - \frac{1}{2}\nab\bcdot \mbs a.
\end{align}
For an incompressible fluid with constant density, together with a volume-preserving (isochoric) condition on the large-scale velocity component, we retrieve the incompressibility conditions (\ref{incompressibility}). It can be noted also that equation (\ref{mass-conservation-1}) still constitutes a transport equation since it preserves energy. As a matter of fact, it can be shown that the energy intake brought by the small-scale component is exactly compensated by the energy dissipation associated to the diffusion term \citep{Resseguier16a}. 
\subsection{Linear momentum conservation}
The application of the stochastic version of the RTT on the stochastic momentum and the introduction of  the forces acting on the flow enables to derive from the second Newton law  the following Navier-Stokes equations \citep{Memin14}:
\begin{subequations}
\label{sto-NS}
\begin{align}
\label{EqInitialSyst1}
&\Bigl(\partial_t \w\!  +\!  \w{\nab}\transp (\w-\frac{1}{2}\nab\bcdot \mbs a)     \! -\! \frac{1}{2} \sum_{ij=1}^d\partial_{x_i} (a_{ij} \partial_{x_j} \w) \Bigr) \rho\!=\! \rho \mbs g  \! -\! \nab p\!  +\!  \mu \Delta \w,\\
\label{EqInitialSyst2}
&\nab p'_t  =
 -   \rho\w\nab\transp \bsigma \dif\B_t +\mu \Delta (\bsigma \dif\B_t),\\
 \label{EqInitialSyst3}
&{\nab} \bcdot(\bsigma \dif \B_t ) =0, \;\;{\nab} \bcdot\w =0, \;\; \nab\bcdot(\nab\bcdot \mbf a) = 0.
\end{align}
\end{subequations}

In this expression $\mu$ is the dynamic viscosity, $p(\xx,t)$ denotes the large-scale (slow)  pressure contribution  and $p'_t$ is a zero-mean turbulent pressure related  to the small-scale velocity component. 
Similarly to the classical Reynolds decomposition, the dynamics of the resolved component includes an additional stress term that depends here on the resolved velocity component. A correction of the advection velocity  also occurs. 
Both terms depend on the variance tensor which gathers the action of the turbulent fluctuations on the large-scale velocity. 

It can be observed that the large-scale energy evolution is dissipative. This generalizes thus the Boussinesq 1877 assumption, which conjectures a dissipative effect of the turbulent fluctuations on the mean flow.  In the case of a divergence-free isotropic random field (with a constant diagonal variance tensor), this system boils down to an intuitive constant eddy viscosity diffusivity model: 
\begin{equation}
(\partial_t \w\! +\! \w{\nab}\transp \w  \! - \! \gamma \frac{1}{2} \Delta\w   )\rho=
\rho \mbs g \! -\!   \nab p\! +\! \mu \Delta \w, \;\nab\bcdot \w=0,
\end{equation}
where the Laplacian dissipation term is augmented by the random field variance. The use of constant eddy viscosity thus finds  here its justification  as a direct consequence of an isotropic turbulence assumption.  

The subgrid stress tensor involved in our formalism constitutes an anisotropic diffusion whose preferential directions of diffusion are given by the small-scale velocity variance tensor. Setting the diffusion tensor, $\bsigma$, or its associated variance tensor allows us to define directly the subgrid diffusion term and the effective advection. For instance imposing to the small-scale random velocity to live on the iso-density surfaces provides immediately a clear justification of the isopycnal diffusion employed in oceanic circulation models \citep{Memin14}. The specification of the turbulent fluctuations in terms of a stochastic process provides  a means to interpret different subgrid models but also to devise new ones either through {\it a priori}  specifications or  data-driven strategies. 

General invariance properties of the proposed large-scale representation are listed in the appendix \ref{Prop-NSU}. We briefly summarize them here. It is in particular shown that the distribution of the velocity anomaly is in the general case not Gaussian and does not consequently correspond to normal or quasi-normal approximations \citep{Monin-YaglomB,Orszag70}. 
We show that this stochastic representation has remarkable invariance properties; it is Galilean invariant and preserves (in the absence of molecular viscosity) the Euler equations' scale invariance properties.   Otherwise, a useful scaling for the variance tensor is derived from the Kolmogorov-Richardson scaling and a dimensionless number relating the large-scale kinetic energy to a characteristic value of the velocity variance at the resolved scale.  

Interested readers may also consult \citep{Resseguier16a,Resseguier16b,Resseguier16c} for  the derivation of stochastic representation of several geophysical flows. 
%
%
%
%
%
%
%
%
%
%
%
%
\section{Numerical simulation and assessment}\label{Experiments}
In this section we assessed, through  numerical simulations, the performances of  the proposed large-scale dynamics for different variance tensor models. Those simple models have been defined from local  statistics of the resolved component and  compared to the classical Smagorinsky subgrid model associated with a dynamical procedure \citep{Germano91,Lilly92}. Those numerical experiments have been performed on the Taylor-Green flow \citep{Taylor-Green}.  
\subsection{Taylor-Green vortex flow simulation}
Taylor-Green vortex flow is a critical test for numerical schemes, as both the convective term and viscous term play important roles. Due to the energy cascade generated by the convective term, the flow becomes rapidly turbulent with the creation of small-scale structures up to a dissipation peak. This stage is followed by a decay phase similar to a decaying homogeneous turbulence. As a consequence, a precise and high-order representation of the viscous and convective terms is needed to get an accurate numerical simulation. This flow is considered as a  prototypical system to study the production of small-scale eddies due to vorticity increase and vortex stretching mechanism \citep{Brachet83,Orszag74,Taylor-Green}.

 In Cartesian coordinates, this flow is defined by the following initial conditions:
\begin{eqnarray*}
&&u(x,y,z)= U_0\sin(\frac{x}{L_0})\cos(\frac{y}{L_0})\cos(\frac{z}{L_0}),
\\
&&v(x,y,z)=-U_0\cos(\frac{x}{L_0})\sin(\frac{y}{L_0})\cos(\frac{z}{L_0}),
\\
&&w(w,y,z)=0,
\end{eqnarray*}
and 
\begin{eqnarray*}
p(x,y,z)= p_0+\frac{\rho U_0^2}{16}\left(\cos(\frac{2x}{L_0})+\cos(\frac{2y}{L_0})+\cos(\frac{2z}{L_0})+2\right).
\end{eqnarray*}
The computation domain is defined as a cubic box with periodic boundary conditions on all the faces. The length of the domain is set to $2\pi$ in each of the axis direction, which gives a characteristic length $L_0=2\pi$ and a Reynolds number $Re=U_0L_0/\nu$.
In the literature, several high-order numerical methods have been tested for the direct numerical simulation (DNS) of the Taylor-Green vortex flow, see \citep{Chapelier13,Rees11} and references therein. In this study we used a discrete scheme built on a divergence-free  wavelet basis \citep{Kadri-IJCV-13,Kadri13}. This scheme presents several computational advantages. First of all, it guarantees a divergence-free solution in the physical domain  with a good numerical complexity. Besides, as the spatial filters considered corresponds to a multi-resolution projection, two successive filtering operations can be switched together. This property reveals useful within the Germano dynamic strategy enabling to estimate the subgrid tensor weight factor. This numerical scheme achieves similar performances to a pseudo-spectral method.  
\subsection{Analysis criterion}
The different numerical simulations performed are mainly assessed and compared according to the evolution along time of the following  criterion:
\begin{itemize}
\item The mean kinetic energy (KE)
\begin{equation*}
{\cal E}(t)=\frac{1}{2|\Omega |}\int_\Omega\w\bcdot\w \dif \x.
\end{equation*}
\item The mean kinetic energy dissipation rate
\begin{equation}
\epsilon(t)=\frac{2\nu}{|\Omega |}\int_\Omega{\mbs S}\boldsymbol{:}{\mbs S}\dif\x,
\end{equation}
\end{itemize}
where ${\mbs S}$ is the rate of strain tensor: $S_{ij}=\frac{1}{2}\left(\partial_{x_i}w_j+\partial_{x_j}w_i\right)$. The mean kinetic energy is linked to the dissipation rate by:
\begin{equation}
\epsilon(t)=-\frac{d{\cal E}}{dt}.
\end{equation}
To clearly separate those two forms of the energy dissipation we will denote $\epsilon_{\cal E}(t)$ the expression computed from the kinetic energy differentiation and $\epsilon_{S} (t)$ the dissipation computed from the rate of strain norm. 

Besides, in the sequel, all those averaged quantities are computed in their dimensionless form:  
\begin{equation*}
t:=\frac{tU_0}{L_0}=t_c,~~{\cal E}(t):=\frac{{\cal E}(t)}{U_0^2}~~\text{and}~~\epsilon(t):=\frac{\epsilon(t)L_0}{U_0^3},
\end{equation*}
where $t_c$ denotes the convective time. The temporal evolution of those mean energy quantities enable to monitor the quality of the solution over time and to assess the accuracy of the discrete scheme used for the velocity gradients. In addition to those criterion, we will plot the energy spectrum of the resolved velocity at several distinct times.
\subsection{Direct Numerical Simulation of the Taylor-Green flow}
We evaluated first the ability  of the divergence-free wavelet based method  to reproduce faithfully the main features of the Taylor-Green flow. For this purpose, two different direct numerical simulations have been conducted. 

\begin{figure}
\centering
\hspace*{-1cm}
\includegraphics[width=0.55\textwidth]{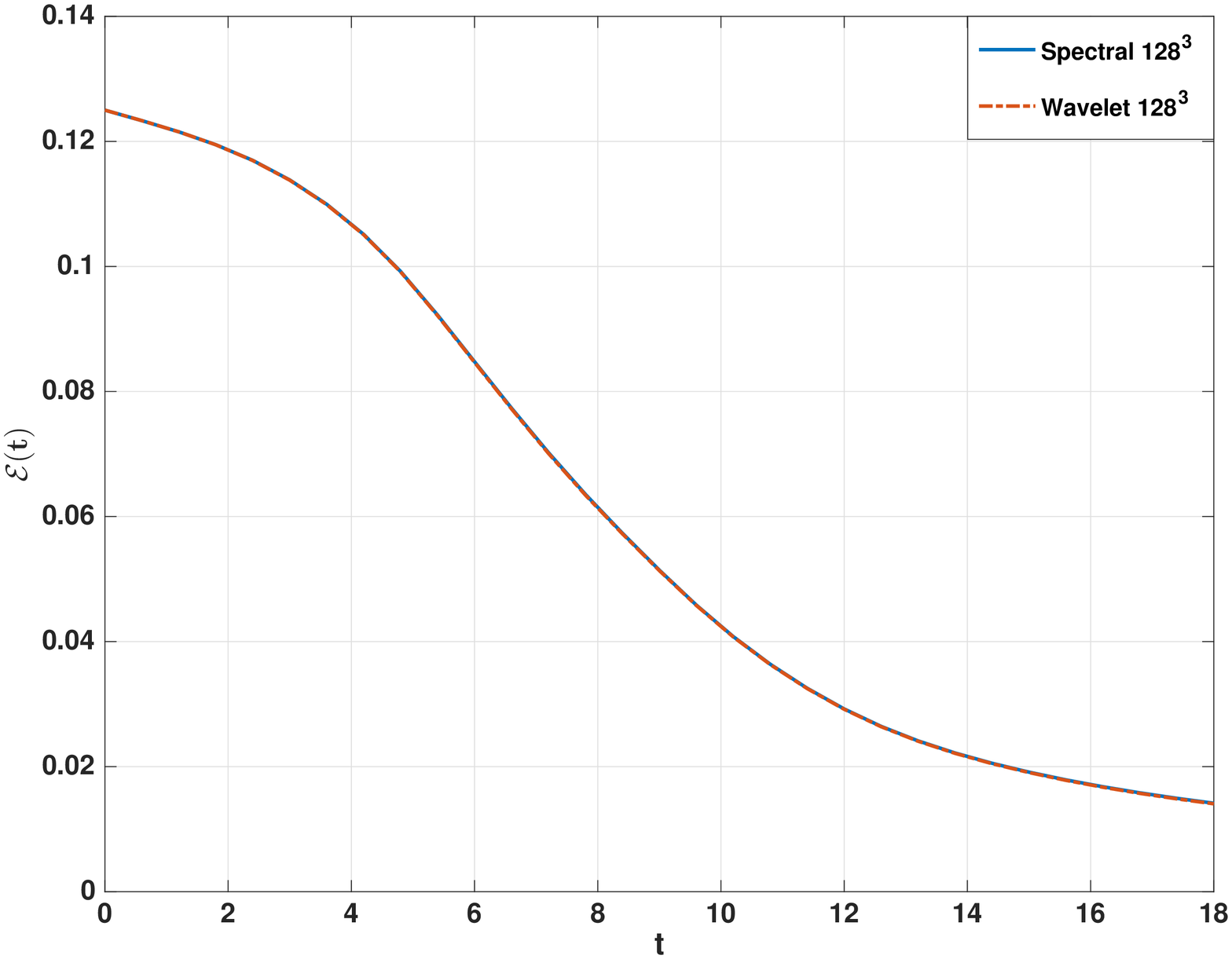}\includegraphics[width=0.55\textwidth]{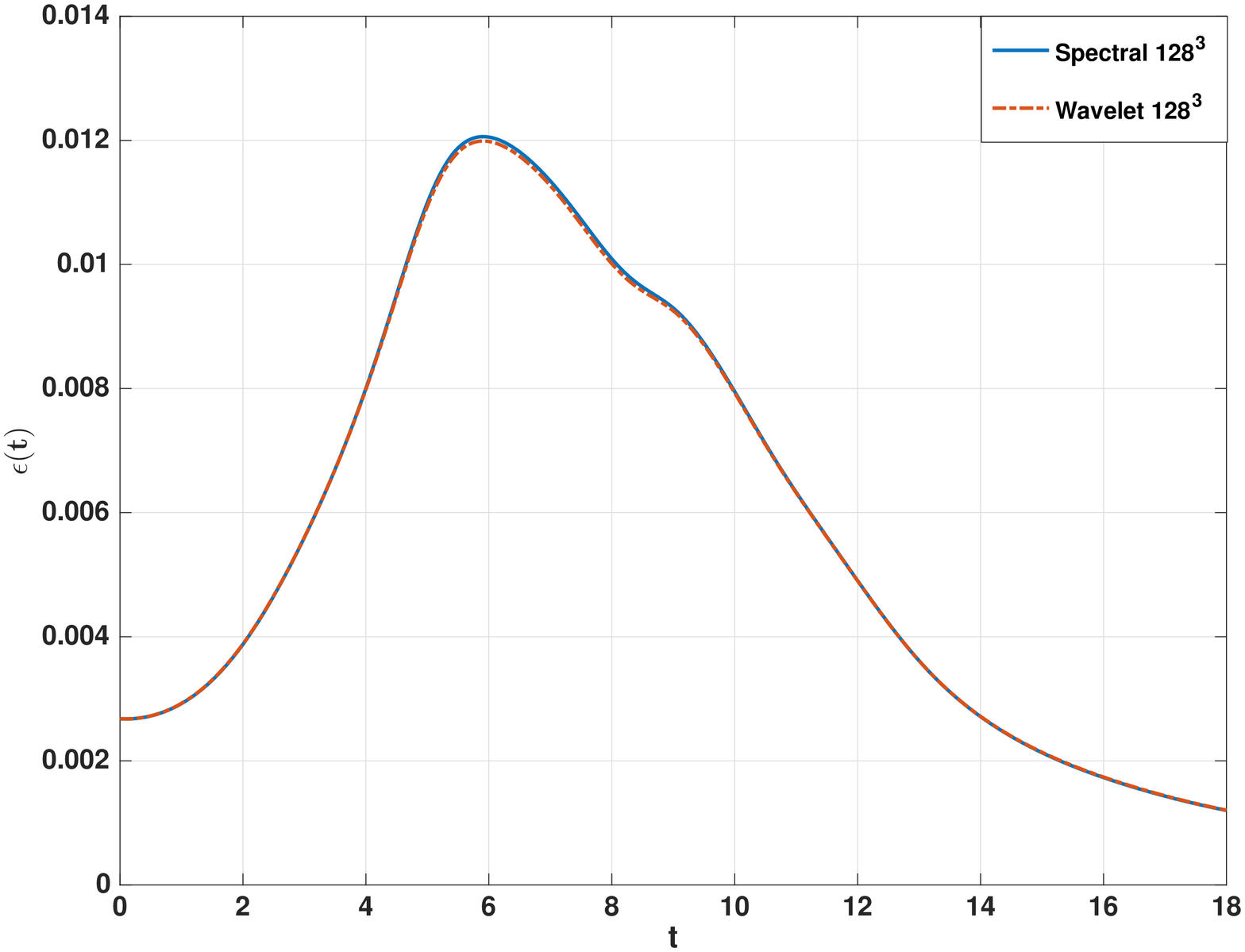}
\caption{Evolution of the dimensionless energy ${\cal E}$ (left) and its dissipation rate $\epsilon_{S}$ (right) as a function of the dimensionless time. Comparison between the divergence-free wavelet method and the Fourier pseudo-spectral method for $\Rey=280$.}
\label{EnergyFig}
\end{figure}

The first experiment concerns a simulation at a low Reynolds number: $\Rey=280$. For this case the divergence-free wavelet has been run on a regular grid of $128^3$ points. For comparison purpose  we performed a classical pseudo-spectral simulation with $128^3$ Fourier modes. Let us note that at this low Reynolds number ($\Rey=280$), only $64^3$ Fourier modes are required to represent accurately all the hydrodynamics scales in the limit of $\delta x\le \eta$, where $\eta$ is the Kolmogorov scale \citep{Chapelier13}. The time evolution of the mean kinetic energy and the mean dissipation rate obtained for both methods are plotted on the left and right panels of figure \ref{EnergyFig} respectively. As can be observed on those two graphics the solutions superimpose almost perfectly. For both methods the energy dissipation computed from the rate of strain tensor norm, $\epsilon_S$, and from the kinetic energy differentiation, $\epsilon_{\cal E}$, coincide perfectly. We plotted therefore only the rate of strain norm. 

\

\begin{figure}
\centering
\hspace*{-1cm}
\includegraphics[width=0.55\textwidth]{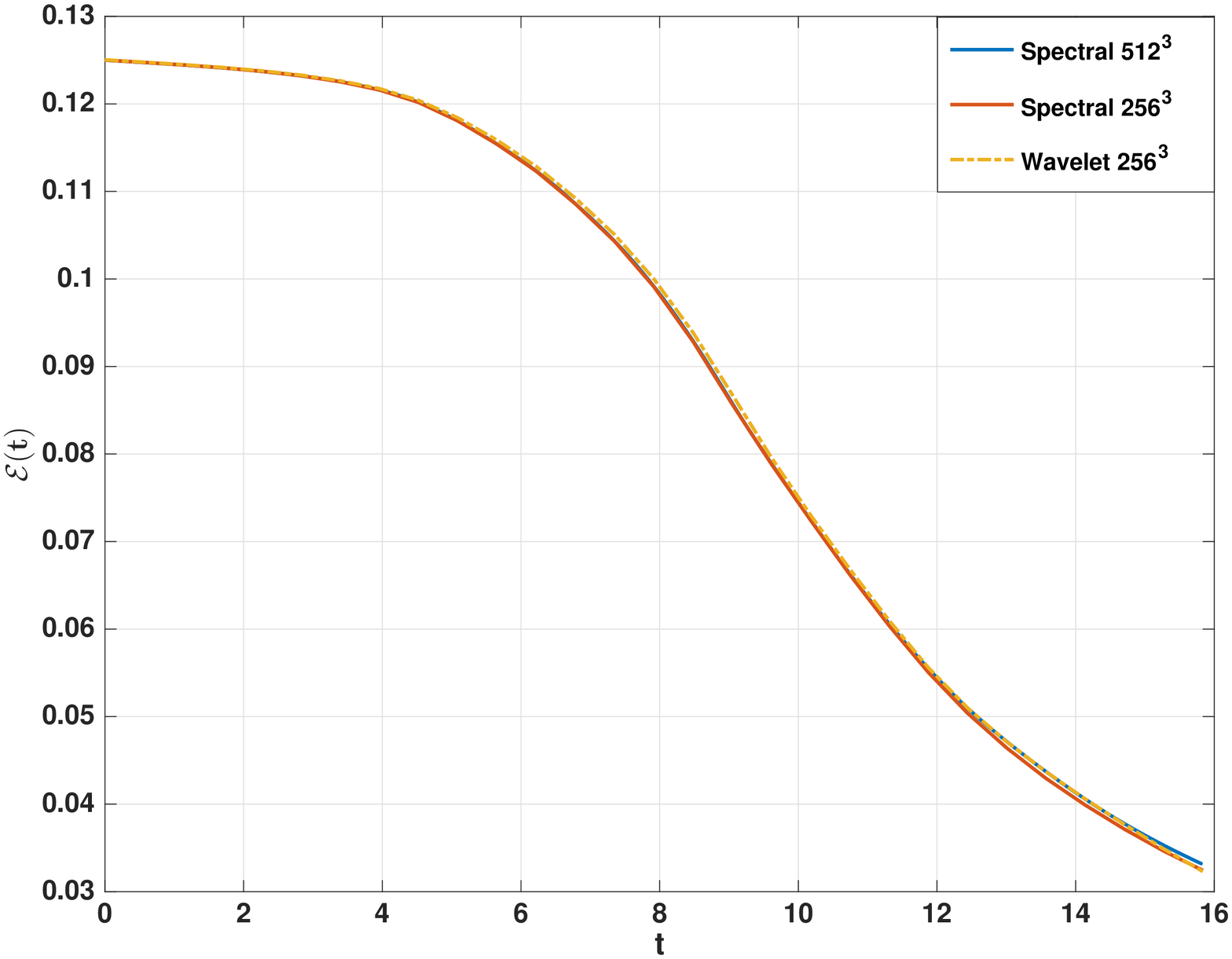}\includegraphics[width=0.55\textwidth]{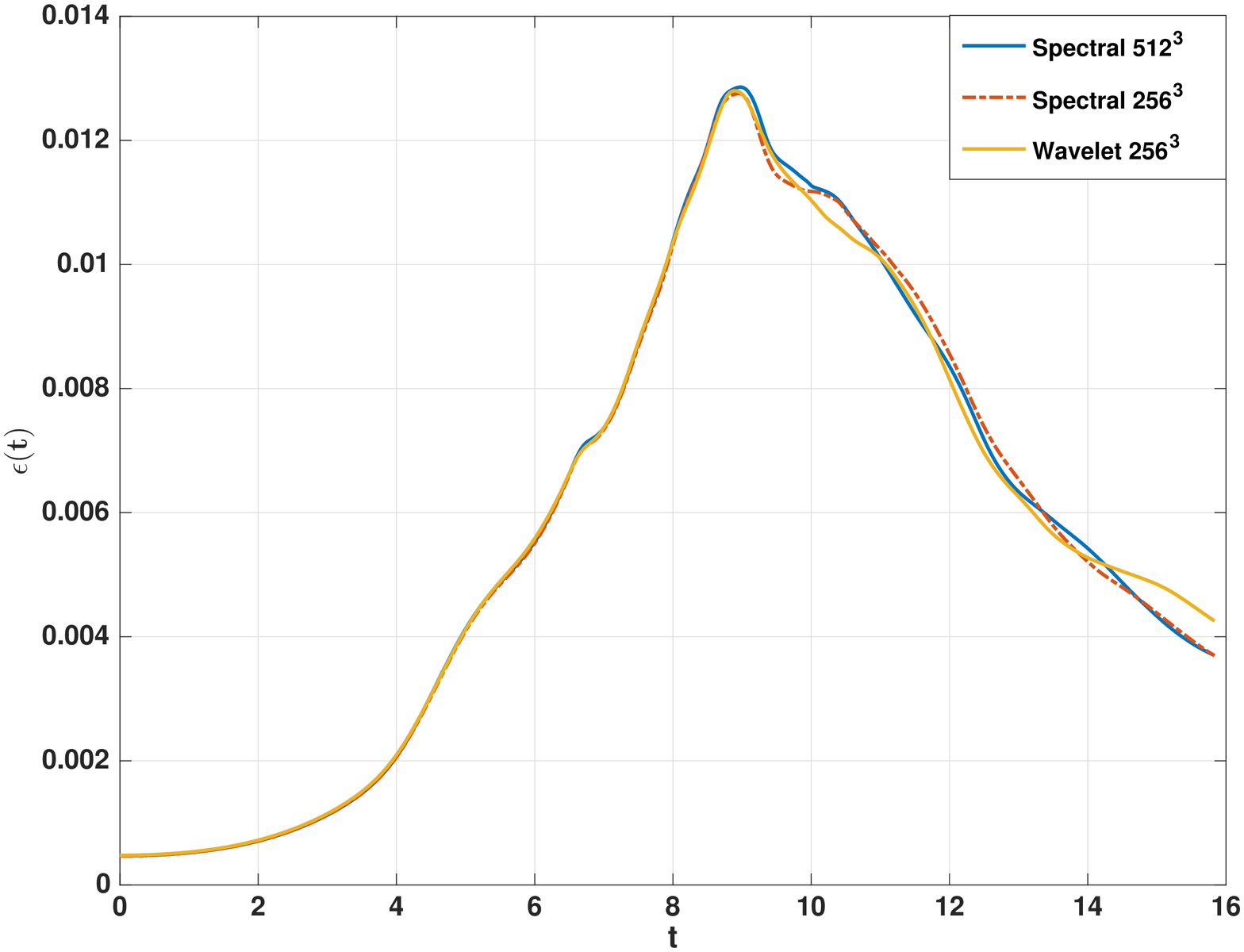}
\caption{Evolution of the dimensionless energy ${\cal E}$ (left) and its dissipation rate $\epsilon_S$ (right) as a function of the dimensionless time. Comparison between the divergence-free wavelet method and two Fourier pseudo-spectral simulations for $Re=1600$.}
\label{EnergyFig1600}
\end{figure}

In a second experiment, both simulations have been then carried out for a moderate Reynolds number fixed at  $\Rey=1600$. Since the study of \citep{Brachet83}, numerous numerical experiences have been conducted to understand accurately the dynamic of the Taylor-Green vortex flow at this Reynolds number. As mentioned in \citep{Chapelier13, Rees11}, all the scales of the flow are captured with $512^3$ degrees of freedom but $256^3$ degrees of freedom are sufficient to represent its main characteristics. We therefore run the divergence-free wavelet based simulation on a $256^3$ mesh grid.  We also performed a Fourier pseudo-spectral simulation with a dealiasing procedure at the same resolution. On figure \ref{EnergyFig1600} we displayed for both simulations the kinetic energy and dissipation rate time evolutions. On the same figure, for comparison purpose, we also plotted the curves corresponding to a $512^3$ Fourier modes pseudo-spectral solution (with dealiasing) available from \citep{Chapelier13}. As in the previous case the dissipation $\epsilon_S$ matches $\epsilon_{\cal E}$. We  therefore only show the rate of strain tensor norm in the right panel of figure \ref{EnergyFig1600}.
It can be observed that the  divergence-free wavelet simulation is in good agreement with the spectral reference solution, especially before the dissipation peak when the convective phenomena is predominant. Slight discrepancies appear in the decaying phase starting after the dissipation peak ($t_c\approx8.95$).

\

Those experiments show that the divergence-free wavelet based method provides  results of comparable quality to those obtained from pseudo-spectral simulations at identical resolutions. The wavelet simulation performed on the $256^3$ mesh grid will serve as a reference "DNS'' for comparison purpose. We relied on this method to carry out all the numerical simulation performed in this study. The performances of various subgrid modeling  are discussed in the next section.
%
%
%
%
%
%
\subsection{Variance tensor and subgrid modelling}
One of the main advantages of the stochastic formalism we propose lies in the great flexibility of the anisotropic diffusion specification. The variance tensor, $\a$, can be fixed from {\em a priori} shape imposed either directly on the small scale variance or on the  diffusion tensor. In some cases, such knowledge could probably be inferred from the physical approximations considered to constitute the model. Aspect ratio simplifications and/or boundary conditions that are not perfectly known could be used as well to constrain the small-scale velocities to live on specific iso-surface. Another route would consist in specifying the turbulent velocity components from small-scale measurements statistics. Despite all those directions are worth exploring as they open new strategies to design sub-grid tensors, in this study we choose to focus on several simple models of tensor $\a(\x,t)$ to explore the potential of the proposed modeling.
\subsubsection{Empirical specification through local mean}
The first model consists in assuming ergodic assumption to compute the variance tensor from local statistics of the resolved velocity component. The variance tensor is here defined from empirical velocity covariance computed  on a local neighborhood:
\begin{multline*}
a_{ij}(\x,n\delta t):=\\\left(\frac{L}{\eta}\right)^{5/3}\delta  t \langle ( w_i({\boldsymbol y},n\delta t)-\mu_i (\x,n\delta t) ) ( w_j({\boldsymbol y},n\delta t)-\mu_j(\x,n\delta t)) \rangle_{{\boldsymbol y}\in{\cal W}(\boldsymbol x,n\delta t)},
\end{multline*}
where $\mu_i(\x,n\delta t)$ denotes the empirical mean on a spatial or temporal window ${\cal W}(\x,n\delta t)$, $L$ is the spatial scale considered for the simulation, which corresponds to a mesh grid composed of $2^{3L}$ grid points, and $\eta$ is the Kolmogorov scale. The empirical averaging, $\langle .\rangle$, is computed either spatially over a small ($3\times3\times3$) window centered around point $\x$ or temporally, at point $\x$, over the time interval $[(n-2)\delta t,n\delta t]$. In the following, they are referred to as the spatial and temporal local covariances respectively. Both models are weighted by the variance tensor scale ratio derived in section \ref{Section-Scaling}.
\subsubsection{Optimal specification through scale similarity}
The second model is defined from two successive filtering of the resolved velocity component and a scale similarity assumption. 
The filtering is defined through the associated wavelet multi-scale projector. More precisely, since the velocity $\w(\x,t)$ can be decomposed as:
%
%
%
%
\begin{equation}\label{DefApproxW2}
\w(\x,t)={\cal P}_{ {\boldsymbol \ell}}[\w(\x,t)]+\sum_{|\boldsymbol j| \ge |{\boldsymbol \ell}|}{\cal Q}_{\boldsymbol j}[\w(\x,t)],~~{\cal Q}_{\boldsymbol j}={\cal P}_{\boldsymbol j+1}-{\cal P}_{\boldsymbol j},
\end{equation}
with ${\cal P}_{\boldsymbol j}$ the projector onto the scaling functions basis \citep{Deriaz06, Kadri13}, the resolved filtered velocity $\w_{\boldsymbol \ell}(\x,t)$ is defined by:
\begin{equation}\label{DefApproxW3}
\w_{\boldsymbol \ell}(\x,t):={\cal P}_{{\boldsymbol \ell}}[\w(\x,t)].
\end{equation}
The second, so called ``test'', filtering is here defined by the projection of $\w_{{\boldsymbol \ell}}(\x,t)$ onto resolution ${{\boldsymbol \ell}'}={\boldsymbol \ell}-1$, which is the immediate coarser resolution following the simulation scale:
\begin{equation}
\w_{\boldsymbol \ell'}(\x,t):= \widetilde{\w_{\boldsymbol \ell}(\x,t)}= {\cal P}_{{\boldsymbol \ell^\prime}}[\w(\x,t)].
\end{equation}
The stochastic Navier-Stokes equations  (\ref{EqInitialSyst1}) rewritten respectively for $\w_{{\boldsymbol \ell}}$ and $\w_{{\boldsymbol \ell'}}$ reads:
\begin{equation}
\label{momentum-test-level1}
\partial_t \w_{\boldsymbol \ell} + \w_{\boldsymbol \ell}{\nab}\transp \w_{\boldsymbol \ell}  - \frac{1}{2}\sum_{i,j=1}^d\partial_{x_i}\partial_{x_j} (a^{\boldsymbol \ell}_{ij}\w_{\boldsymbol \ell}) = -\frac{1}{\rho} {\nab p_{\boldsymbol \ell}} + \nu \Delta\w_{\boldsymbol \ell}+\f_{\boldsymbol \ell},
\end{equation}
and
\begin{equation}
\label{momentum-test-level2}
\partial_t \w_{\boldsymbol \ell'} + \w_{\boldsymbol \ell'}{\nab}\transp \w_{\boldsymbol \ell'}  - \frac{1}{2}\sum_{i,j=1}^d\partial_{x_i}\partial_{x_j} (a^{\boldsymbol \ell'}_{ij}\w_{\boldsymbol \ell'}) = -\frac{1}{\rho} {\nab p_{\boldsymbol \ell'}} + \nu \Delta\w_{\boldsymbol \ell'}+\f_{\boldsymbol \ell'}.
\end{equation}
Taking the difference of the momentum equations \eqref{momentum-test-level1} and \eqref{momentum-test-level2} for the two subsequent levels ${\boldsymbol \ell}$ and ${\boldsymbol \ell'}$ provides the residual dynamics:
\begin{multline}
\partial_t \bar\w_{\boldsymbol \ell} +\frac{1}{\rho} {\nab \bar p_{\boldsymbol \ell}} - \nu \Delta{\bar \w_{\boldsymbol \ell}} -\bar{\f}_{\boldsymbol \ell}= \\-\Bigl(\w_{\boldsymbol \ell}{\nab}\transp \w_{\boldsymbol \ell}-\w_{\boldsymbol \ell'}{\nab}\transp \w_{\boldsymbol \ell'} + \frac{1}{2}\sum_{i,j=1}^d\partial_{x_i}\partial_{x_j} ({a^{\boldsymbol \ell}_{ij}(\w_{\boldsymbol \ell}-\lambda\w_{\boldsymbol \ell'}})\Bigr),
\end{multline}
where $\bar\w_{\boldsymbol \ell}=\w_{\boldsymbol \ell}-\w_{\boldsymbol \ell}'$, $\bar\f_{\boldsymbol \ell}=\f_{\boldsymbol \ell}-\f_{\boldsymbol \ell}'$ and $\lambda\in\R$. Note that a similarity assumption $\a^{\boldsymbol \ell'}=\lambda \a ^{\boldsymbol \ell} $ has been imposed for the variance tensor. It can be noticed that, if the projector ${\cal P}_{\boldsymbol \ell}$ commutes with differentiation, due to the filtering projection property, the Stokes equation in the left-hand side cancels after filtering. Then, instead of solving the right-hand expression at level ${\boldsymbol \ell'}$ and then projecting back at a finer level the estimated variance tensor, we rather propose to solve it at scale, $\boldsymbol \ell$, in a least squares sense. Introducing the variance tensor incompressibility constraint we seek the minimizer of the following nonlinear functional:
\begin{multline*}
{\cal J(\a,\lambda)}=\\\frac{1}{2}\bigl\| \w_{ {\boldsymbol \ell}}{\nab}\transp (\w_{ {\boldsymbol \ell}}-\frac{1}{2}\nab\bcdot \mbs a^{\boldsymbol \ell})- \w_{ {\boldsymbol \ell'}}{\nab}\transp (\w_{ {\boldsymbol \ell'}}-\frac{\lambda}{2}\nab\bcdot \a^{\boldsymbol \ell})-\frac{1}{2}\sum_{i,j=1}^d\partial_{x_i}[a^{\boldsymbol \ell}_{ij}\partial_{x_j}(\w_{\boldsymbol \ell}-\lambda\w_{{\boldsymbol \ell}'})\bigr\|^2_{L^2(\R^d)},
\end{multline*}
with the positivity constraint:
\begin{equation*}
\sum_{i,j=1}^da^{\boldsymbol \ell}_{ij}(\x,t)\xi_i\xi_j>\alpha\|\xi\|^2,~~\forall\xi\in\R^d,~\alpha>0,
\end{equation*}

\

In practice the minimization of the functional ${\cal J}$ has been carried out  using a quasi-newton method combined with the Broyden-Fletcher-Goldfarb-Shanno (BFGS) method \cite {Nocedal99} to approximate the Hessian matrix. The variance tensor has been assumed to be diagonal at all points. In order to impose the positivity constraint, instead of computing $\a(\x,t)$ we preferred to compute its square root $\sqrt{\a(\x,t)}$. For details on the computation of the gradient of ${\cal J}$, we refer to  appendix \ref{GradientComputation}. 
\subsubsection{Smagorinsky subgrid model}
The third model evaluated corresponds to the  classical Smagorinsky eddy viscosity formulation,
\begin{equation*}
\nu_t= (C_s \delta\xx_{\boldsymbol \ell})^2|\S|,~~|\S|= \left(2\sum_{i,j=1}^d  S^2_{ij}\right)^{\frac{1}{2}},
\end{equation*} coupled with the Germano estimation procedure  \citep{Germano91,Lilly92} to fix dynamically the eddy viscosity constant from a least squares estimation \citep{Lilly92} and a filtering of the velocity field at two consecutive scales. The ``test'' filtering is as previously defined by the projection of $\w_{\ell}(\x,t)$ onto the immediate coarser resolution ${\boldsymbol \ell}'={\boldsymbol \ell}-1$. Denoting the filtering operation  $\widetilde{\w_{\boldsymbol \ell}}=\w_{{\boldsymbol \ell}'}(\x,t)$, the constant $C_s(t)$ is given by:
\begin{equation*}
C_s^2(t)=\frac{\langle {\cal M}_{ij}({\cal L}_{ij}-\frac{1}{3}{\cal L}_{kk}\delta_{ij})\rangle}{\langle{\cal M}_{kl}{\cal M}_{kl}\rangle},
\end{equation*}
with 
\begin{equation*}
{\cal L}_{ij}=\widetilde{w_i w_j}-\widetilde{w}_i\widetilde{w}_j~~~\text{and}~~~{\cal M}_{ij}=2(\delta\x_{\boldsymbol \ell})^2\widetilde{|\S|S_{ij}}-2(\delta\x_{\boldsymbol \ell'})^2|\widetilde{\S}|\widetilde{S}_{ij},
\end{equation*}
and $\langle .\rangle $ denotes a spatial averaging.
\subsection{Large-scale simulation numerical results}
\begin{figure}
\centering
\hspace*{-1cm}
\includegraphics[width=7.5cm,height=6.0cm]{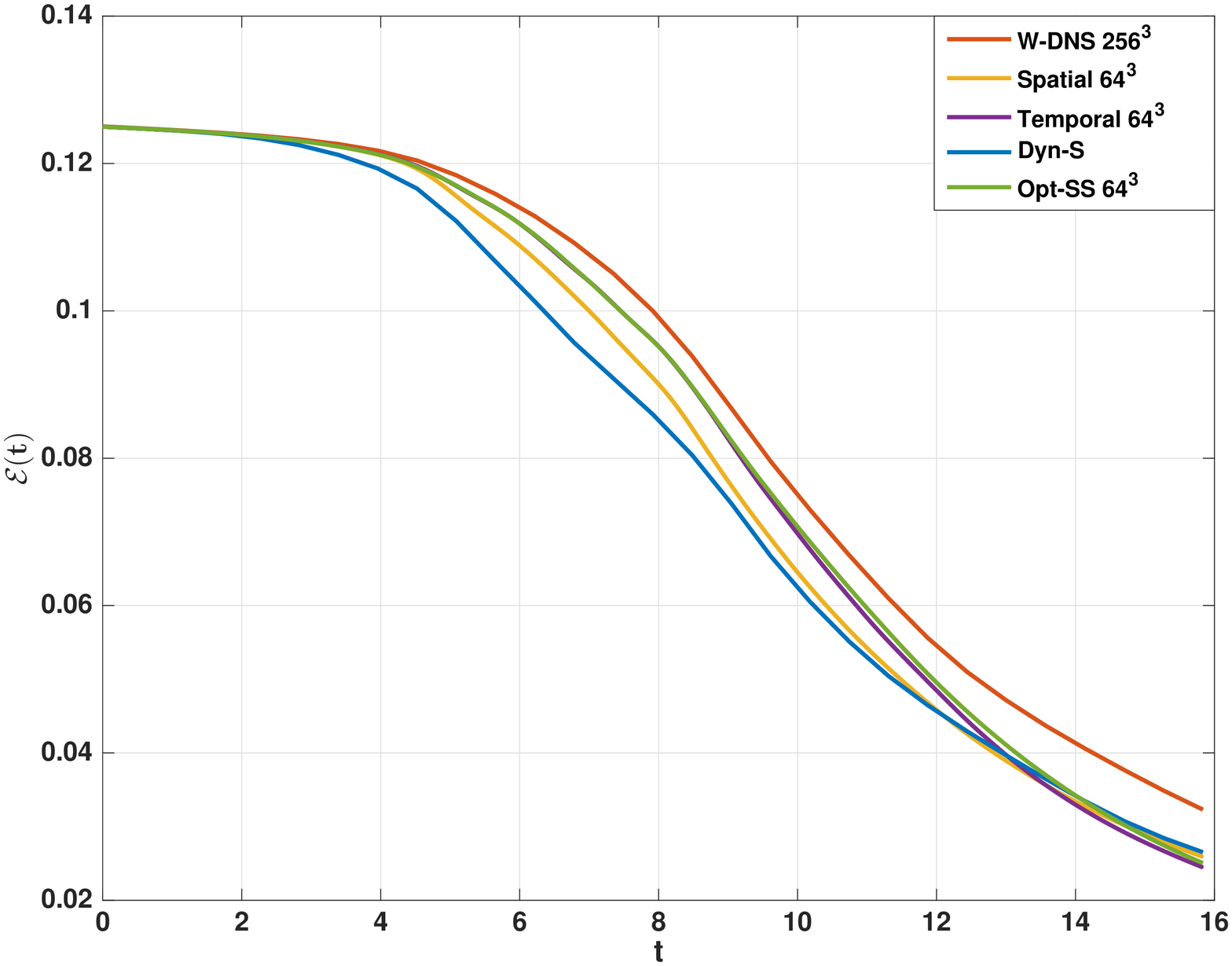}\includegraphics[width=7.5cm,height=6cm]{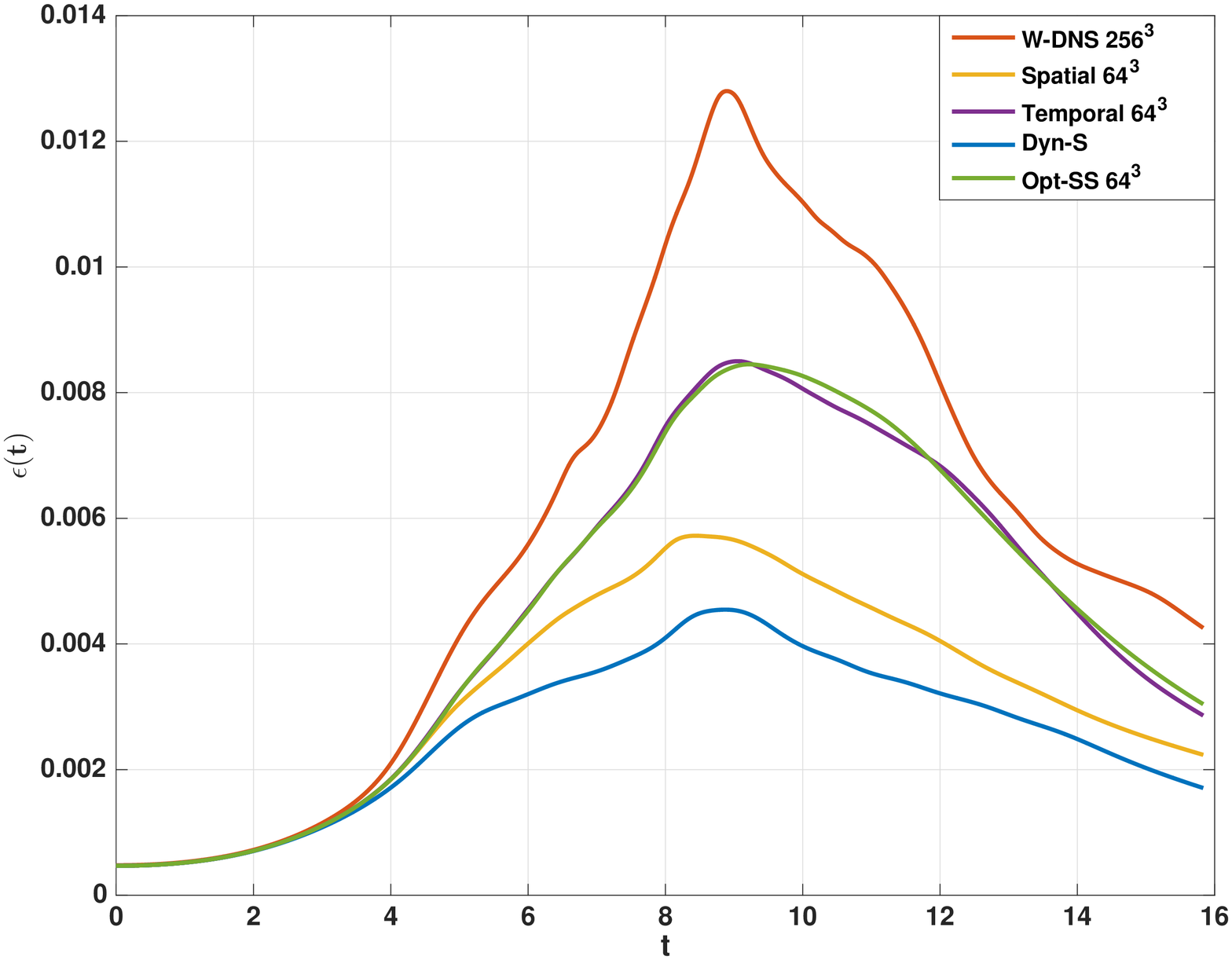}
\caption{Evolution of the dimensionless energy ${\cal E}(t)$ (left) and of the energy dissipation rate $\epsilon_S(t)$ (right) as a function of the dimensionless time. Comparison between the three variance tensor models, the dynamic Smagorinsky model and the DNS wavelet-based solution for the Taylor-Green vortex at $Re=1600$.}
\label{EnergyFigLES}
\end{figure}

The performances of the different sub-grid models have been compared for a $64^3$ mesh grid, which corresponds to a resolution scale $L=6$. Figure \ref{EnergyFigLES} shows for the different models the time evolution of the mean kinetic energy (left panel) and the mean energy dissipation rate (right panel). Those curves can be compared to the reference solution computed on a $256^3$ grid ($L=8$). An important remark can be raised here. The filtered velocity fields do not correspond to the solution of a Navier-Stokes equation started from the large-scale (filtered) initial conditions.  It is thus not relevant to correlate the large-scale solutions with the filtered velocity. It is more meaningful to compare them directly with the DNS. 

\begin{figure}
\centering
\hspace*{-1cm}
\includegraphics[width=7.5cm,height=6.0cm]{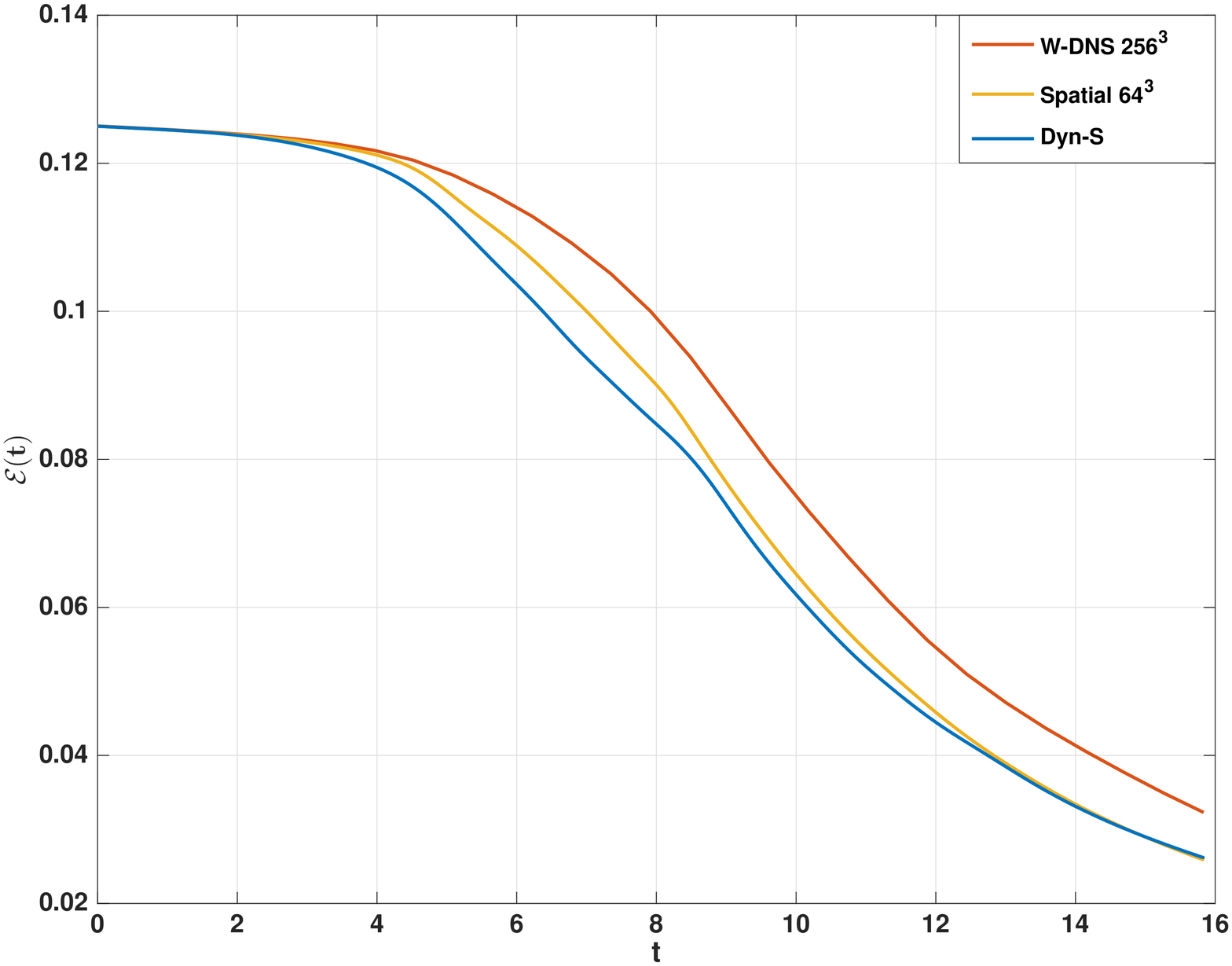}\includegraphics[width=7.5cm,height=6cm]{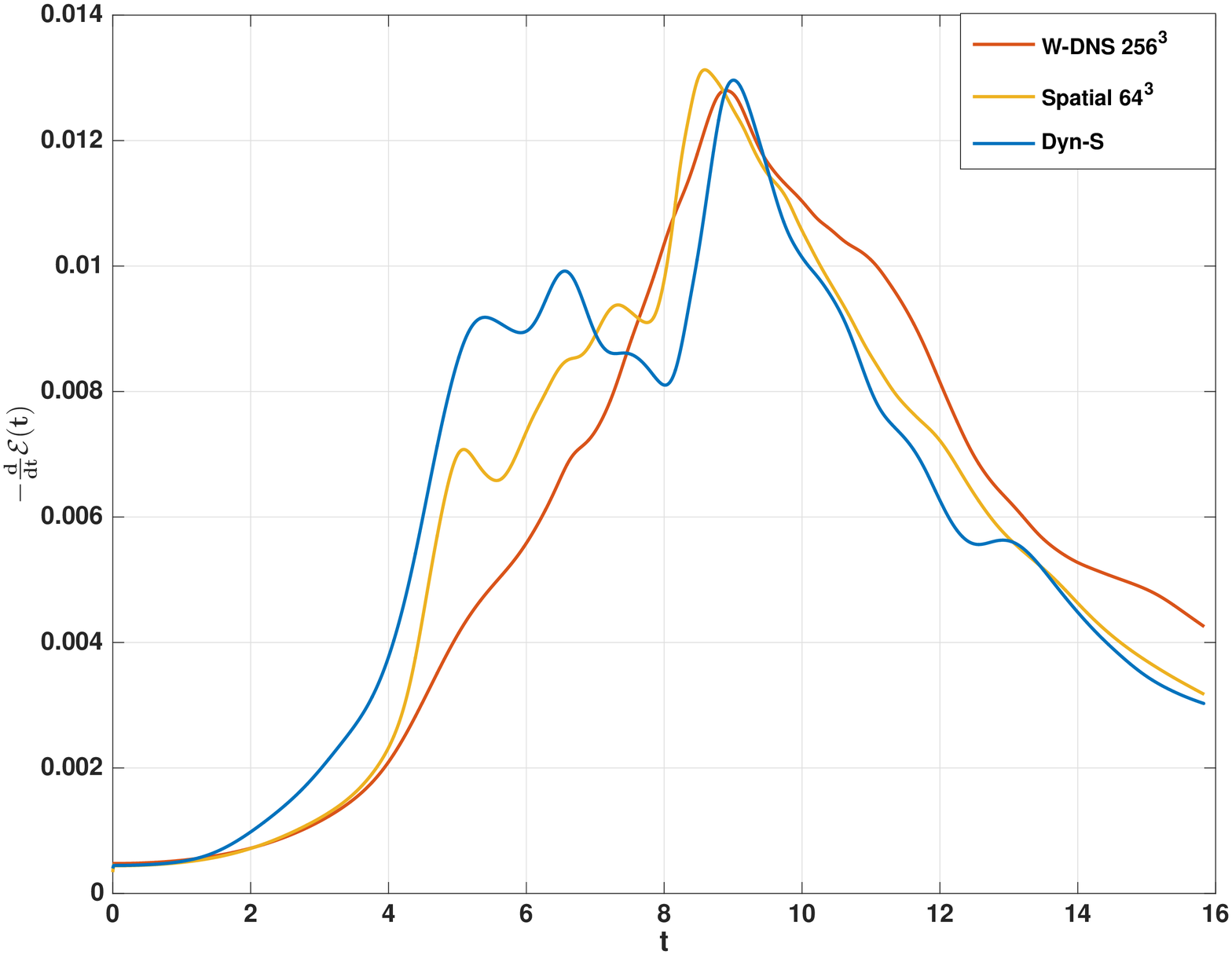}
\caption{Evolution of the dimensionless energy ${\cal E}(t)$ (left) and of the energy dissipation rate $\epsilon_{\cal E }=-\frac{d}{dt}{\cal E}(t)$ (right) as a function of the dimensionless time. Comparison between the spatial covariance model, the dynamic Smagorinsky model  and the DNS wavelet-based solution for the Taylor-Green vortex at $Re=1600$.}
\label{EnergyFigLES2}
\end{figure}

\begin{figure}
\centering
\hspace*{-1cm}
\includegraphics[width=7.5cm,height=6.0cm]{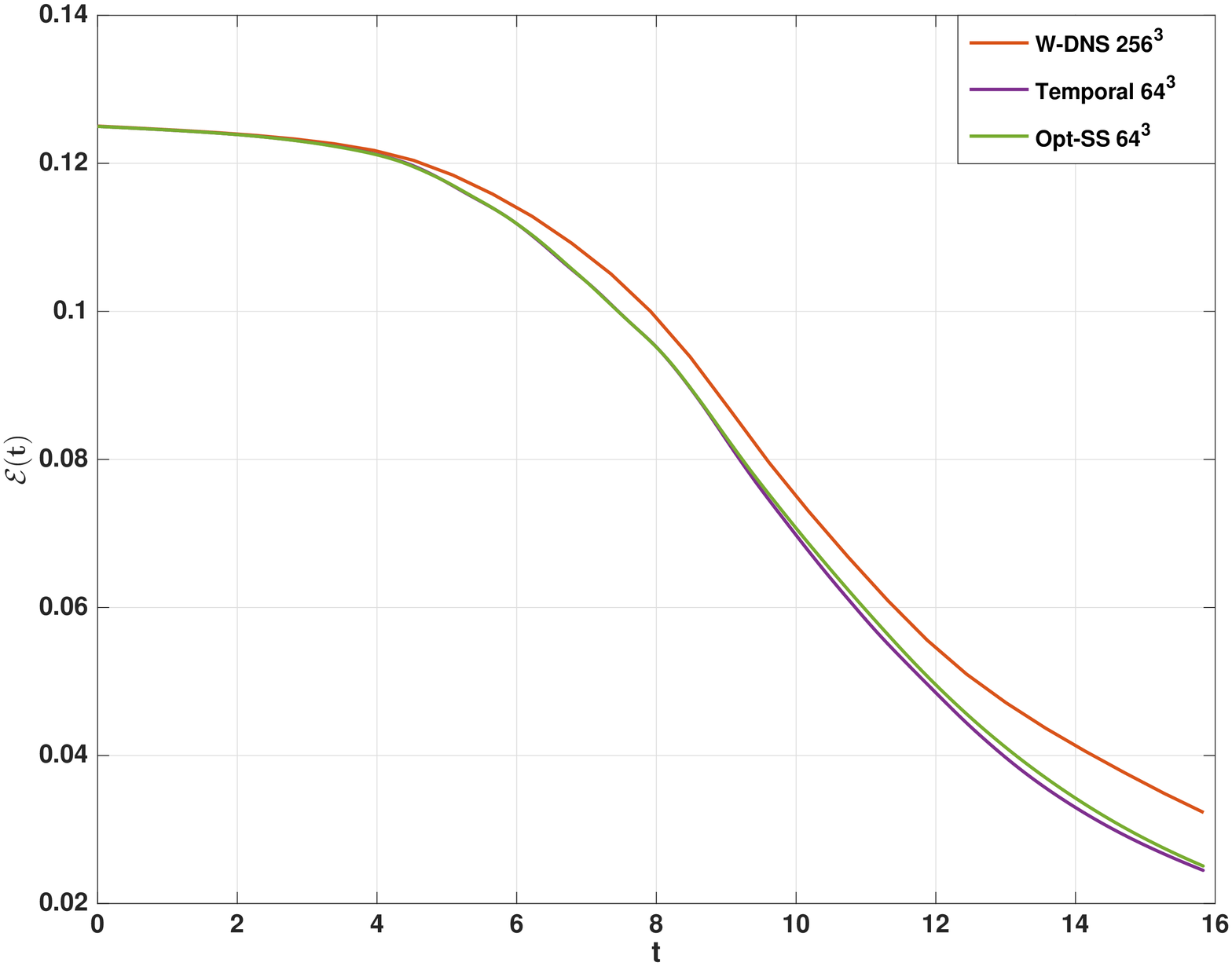}\includegraphics[width=7.5cm,height=6cm]{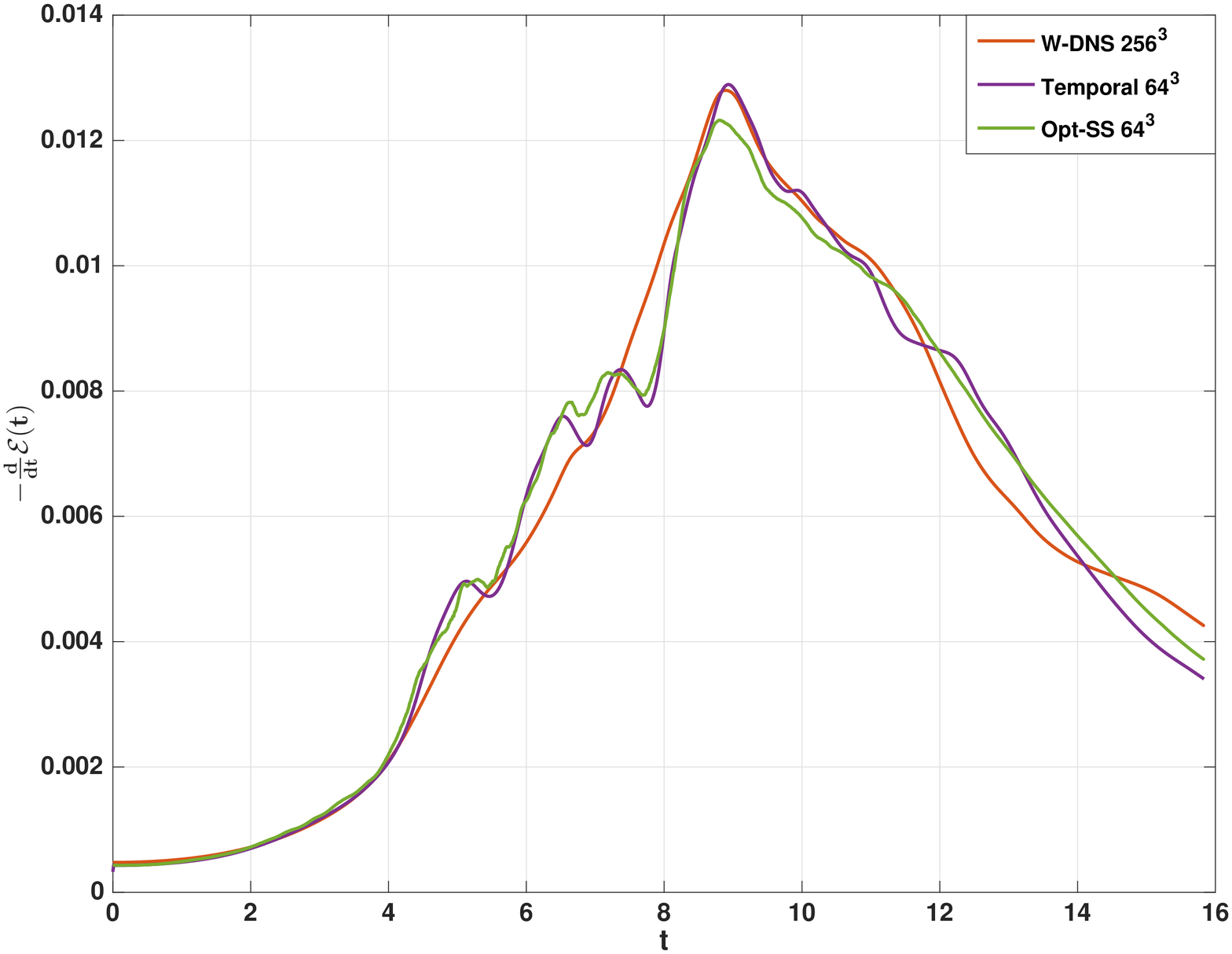}
\caption{Evolution of the dimensionless energy ${\cal E}(t)$ (left) and of the energy dissipation rate  $-\frac{d}{dt}{\cal E}(t)$ (right) as a function of the dimensionless time. Comparison between the temporal local covariance model, the optimal scale similarity model and the DNS wavelet-based solution for the Taylor-Green vortex at $Re=1600$.}
\label{EnergyFigLES3}
\end{figure}

From the rate of strain tensor norm displayed on the right side of figure \ref{EnergyFigLES}, two groups of  subgrid tensor can be recognized. The first one gathers the dynamic Smagorinsky tensor ({\tt Dyn-S}) and the variance tensor defined from the spatial local velocity covariance ({\tt Spatial}). The second group is composed of the optimal variance tensor with a scale similarity assumption ({\tt Opt-SS}) and the variance tensor built from the velocity temporal covariance ({\tt temporal}). For both groups, we displayed on the right-hand side of figure \ref{EnergyFigLES2} and figure \ref{EnergyFigLES3} the energy dissipation rate computed from the kinetic energy rate of change. On the left-hand side we plotted the kinetic energy associated to each model. Those curves can be compared to the DNS reference curve. We observe from those figures that all the large-scale models achieved to reproduce the right amount of kinetic energy and energy dissipation rate. The dynamic Smagorinsky subgrid tensor is the model which provides the lower kinetic energy and the lower rate of strain norm. We note the first group of subgrid tensors exhibits in a general way a too fast dissipation in the first phase of the flow ($t\in [4,9]$), characterized by the domination of the advection mechanism. An undue smoothing of the resolved velocity gradient results from this too high dissipation rate. This is confirmed by looking at the rate of strain norm plotted in figure \ref{EnergyFigLES}. Within this first group, the model build from local spatial velocity covariance  performs  better than the original Smagorinsky tensor. The second group of variance tensor outperforms clearly the results of the first group. Both the temporal covariance model and the optimal variance tensor show strikingly very close results. The optimal variance tensor has a slightly higher dissipation rate at the dissipation peak whereas the temporal covariance model fits almost perfectly the DNS results around the dissipation peak ($t\in [8,10]$). Both models slightly differ from the DNS in the decaying phase. In a general way, we note that all the models have a too slow dissipation rate at the end of the decaying phase ($t>14$). They all smooth too much the velocity gradients in that regime. It must be  outlined that the good behavior of the temporal covariance confirms the relevance of the variance tensor scaling since no dynamics strategy has been performed in that case.

\begin{figure}
\centering
\hspace*{-1cm}
\includegraphics[width=0.55\textwidth]{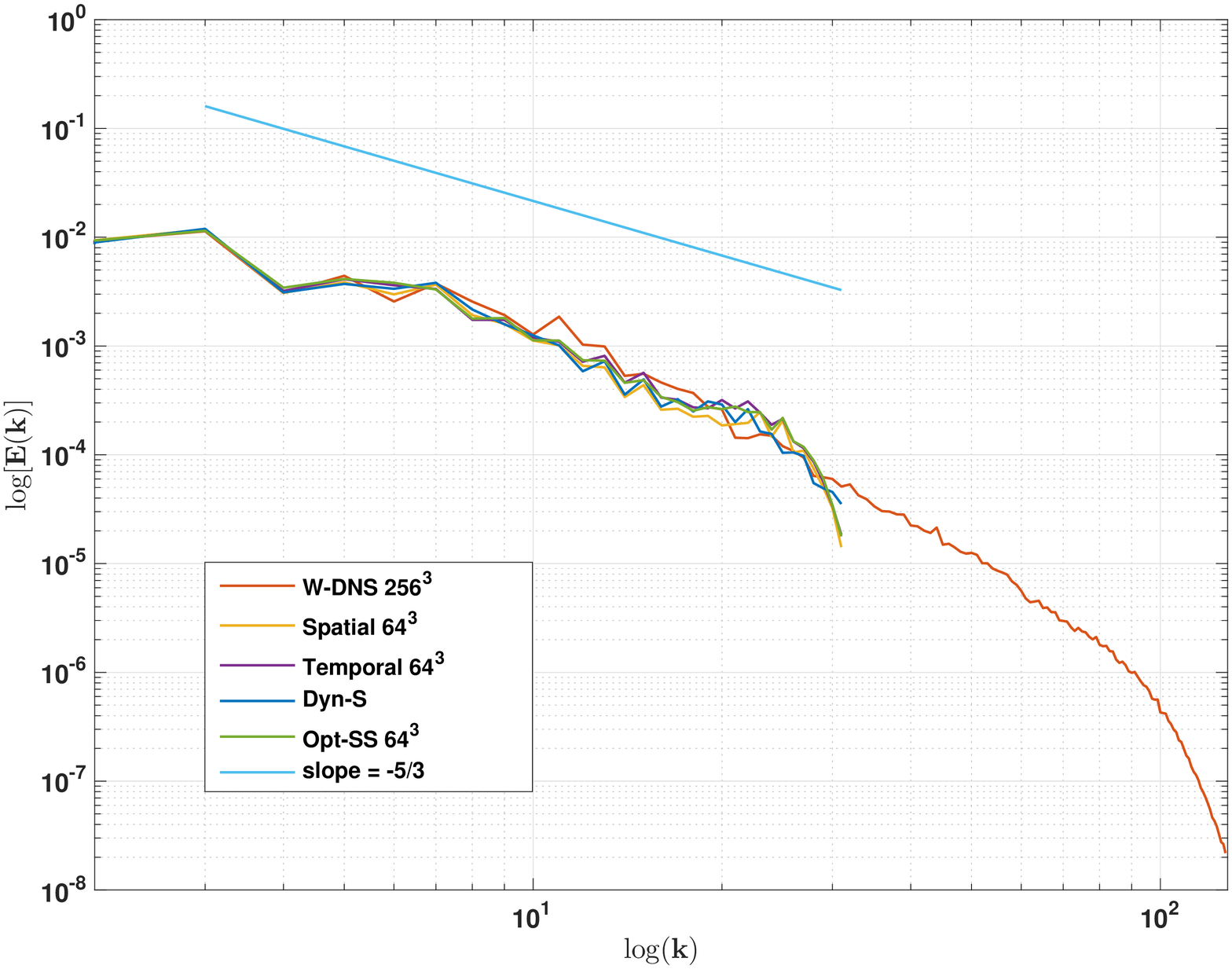}\includegraphics[width=0.55\textwidth]{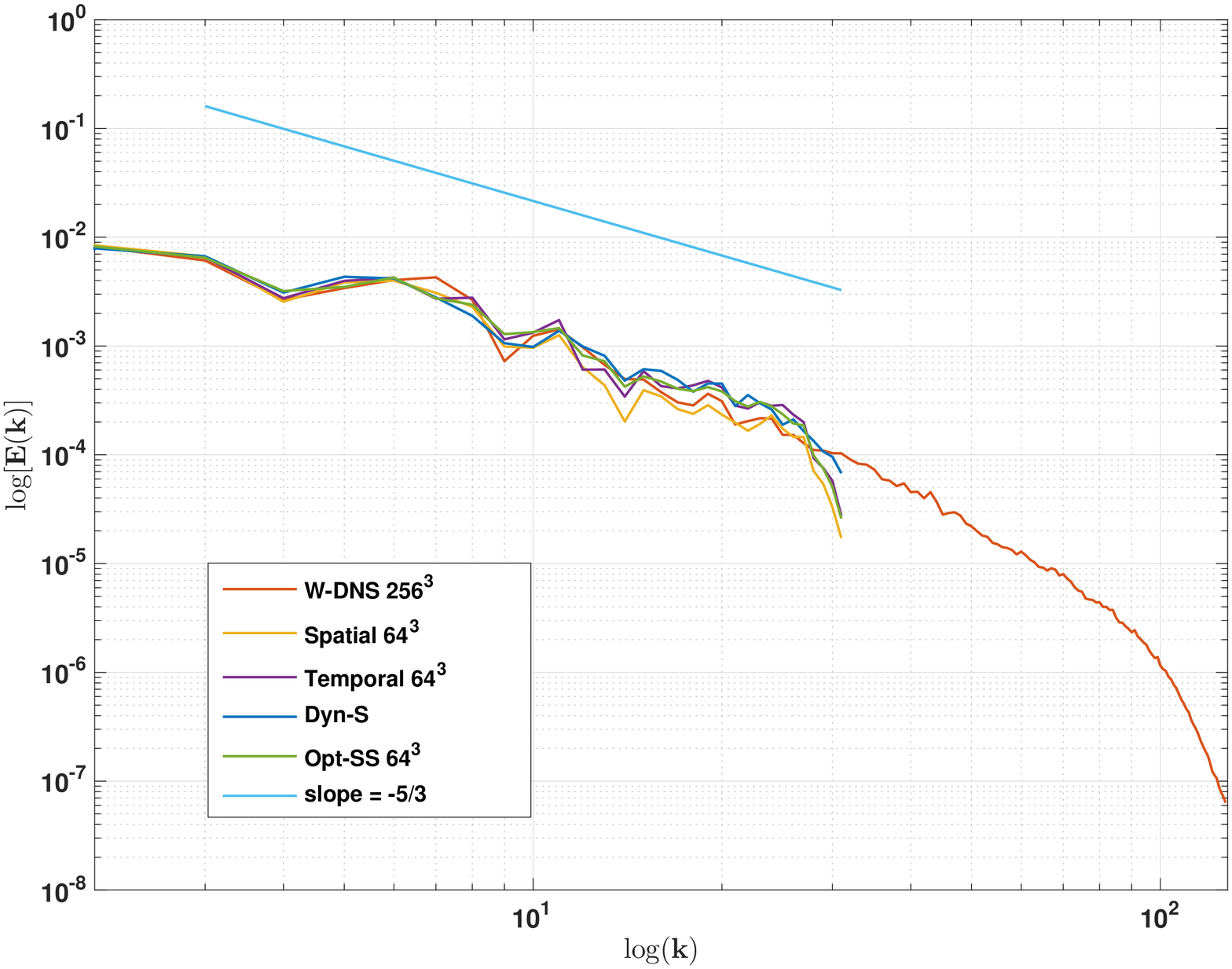}
\caption{Total energy spectrum at $t\approx 7.91$ (left) and $t\approx 9$ (right). Comparison between the three  variance tensor models, the dynamic Smagorinsky model and the DNS solution. Taylor-Green vortex at $Re=1600$.}
\label{SpectreLES1}
\end{figure}
\begin{figure}
\centering
\hspace*{-1cm}
\includegraphics[width=0.55\textwidth]{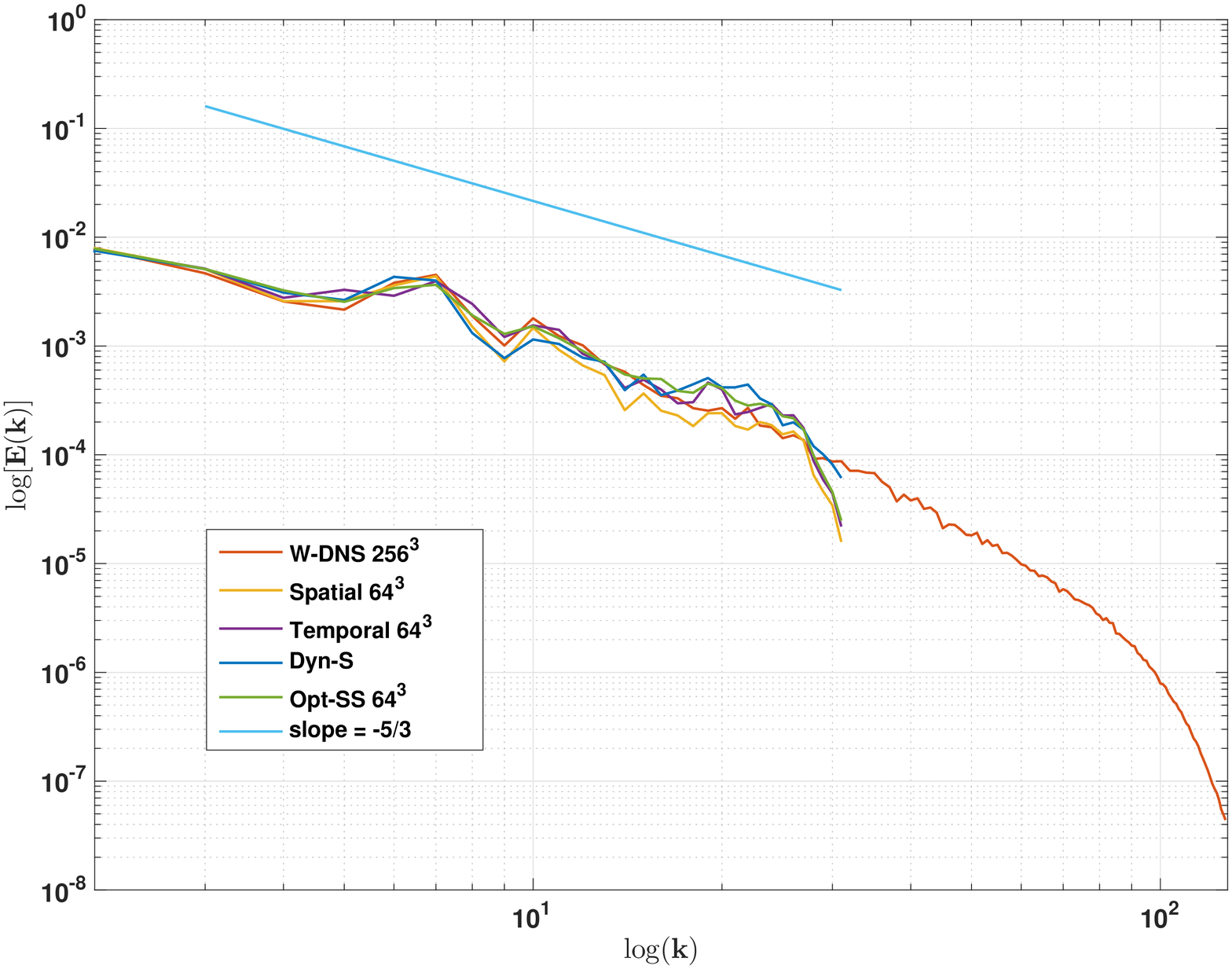}\includegraphics[width=0.55\textwidth]{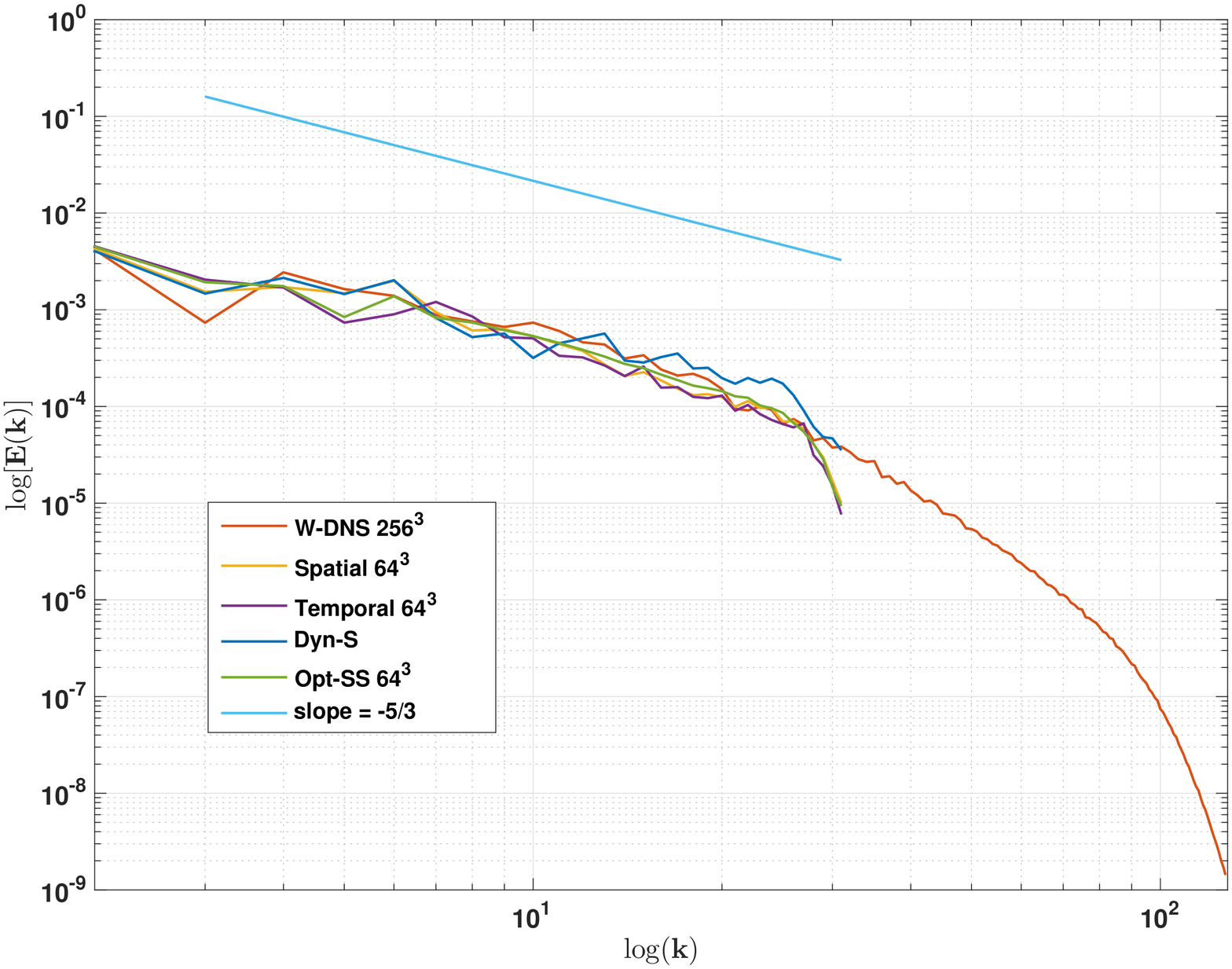}
\caption{Total energy spectrum at $t\approx9.61$ (left) and $t\approx 14.7$ (right). Comparison between the three variance tensor  models, the dynamic Smagorinsky model and the DNS solution. Taylor-Green vortex at $Re=1600$.}
\label{SpectreLES2}
\end{figure}

We display next on figure \ref{SpectreLES1}  and on figure \ref{SpectreLES2}  the energy spectrum associated to the different models  at four different instants. All the models provide satisfying solutions with similar spectrum. Compared to the other models, the dynamic Smagorinsky subgrid stress produces a noticeable energy bump at the cutoff scale. The temporal and optimal variance tensor have spectrum which are in general closer to the DNS spectrum. At the end of the turbulence decay phase ($t>13$) all the models provide close results. 

\begin{figure}
\centering
\hspace*{-1cm}
\includegraphics[width=0.55\textwidth]{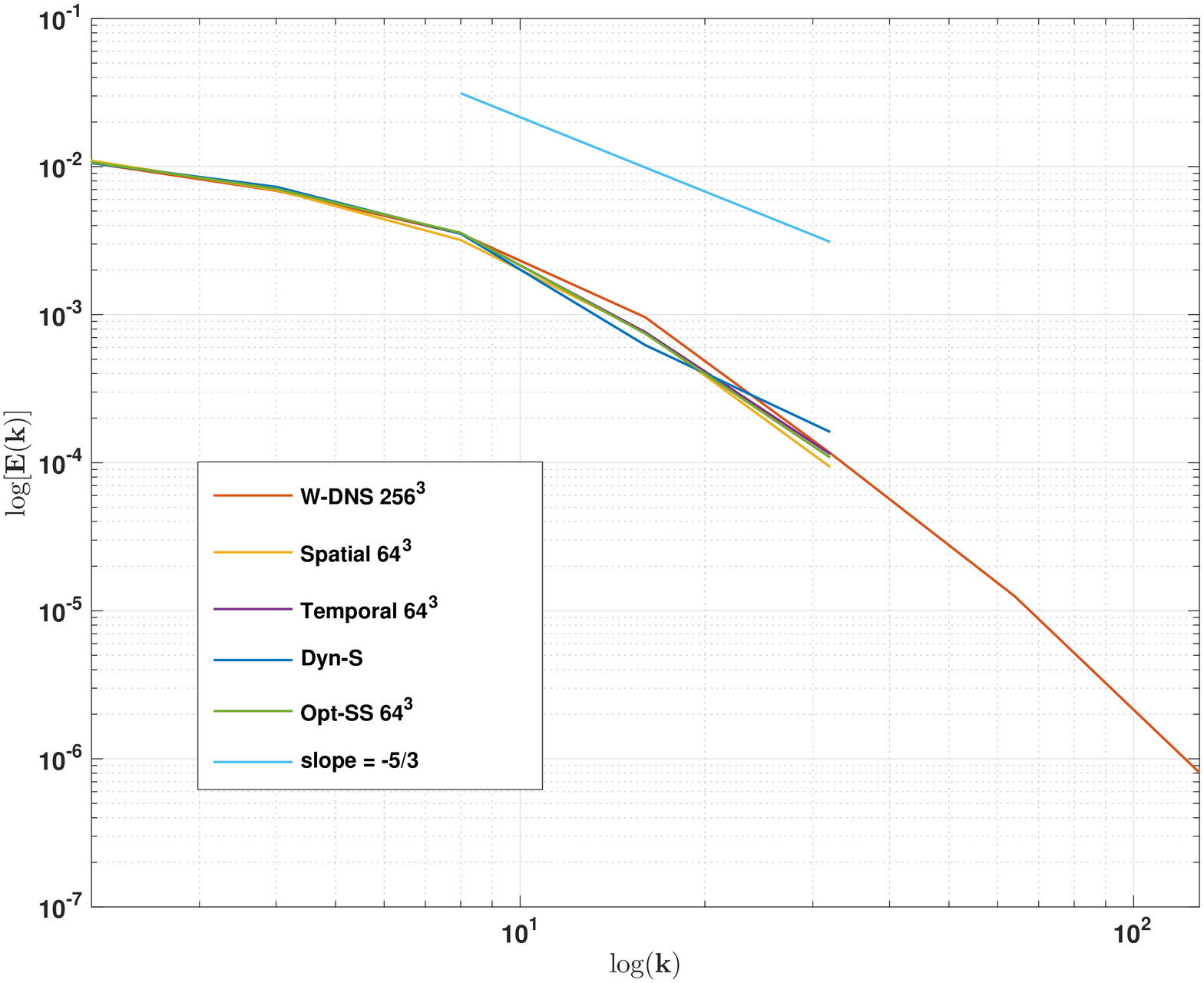}\includegraphics[width=0.55\textwidth]{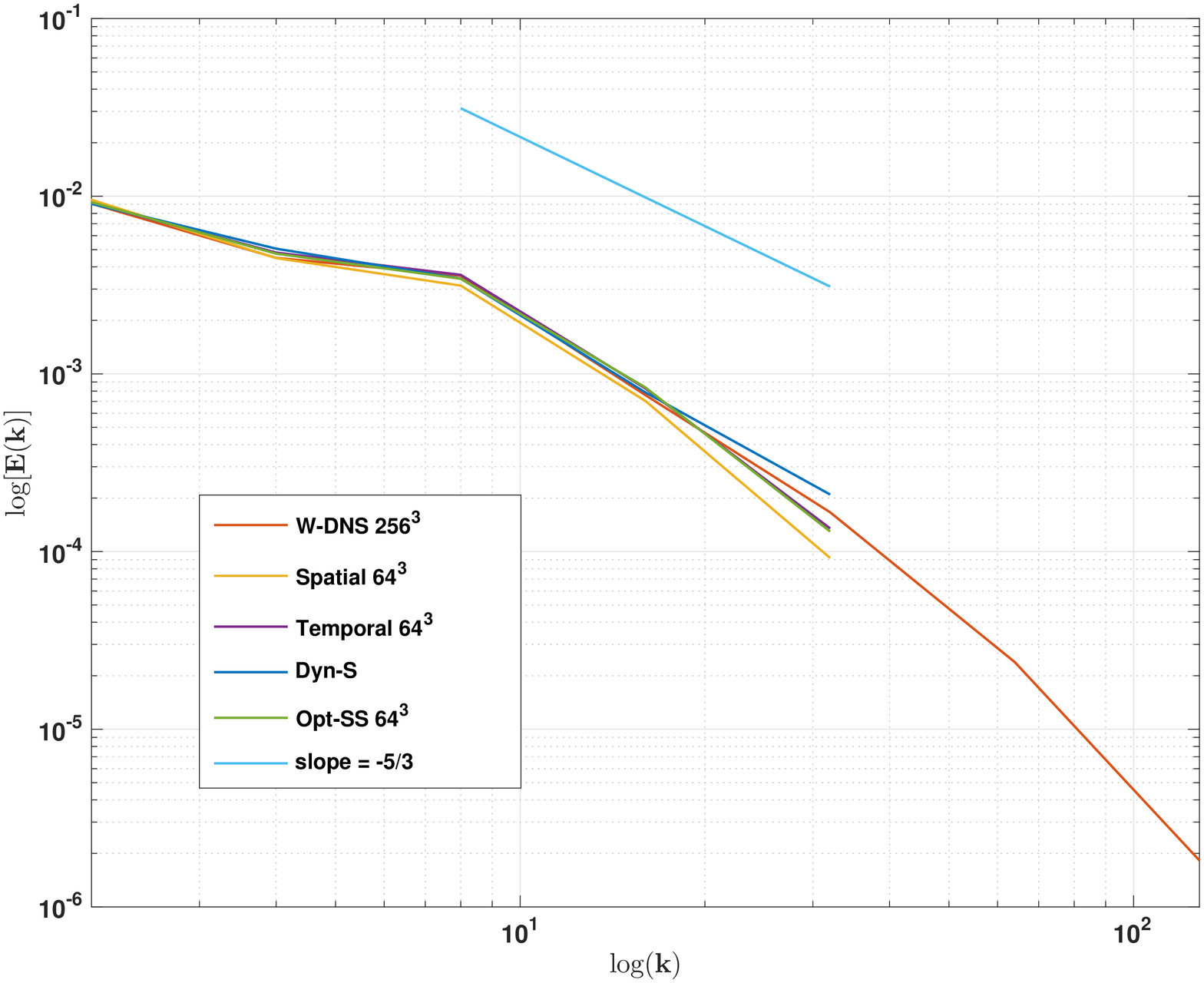}
\caption{Wavelet energy spectrum at $t\approx 7.91$ (left) and $t\approx 9$ (right). Comparison between the three  variance tensor models, the dynamic Smagorinsky model and the DNS solution. Taylor-Green vortex at $Re=1600$.}
\label{WSpectreLES1}
\end{figure}

\begin{figure}
\centering
\hspace*{-1cm}
\includegraphics[width=0.55\textwidth]{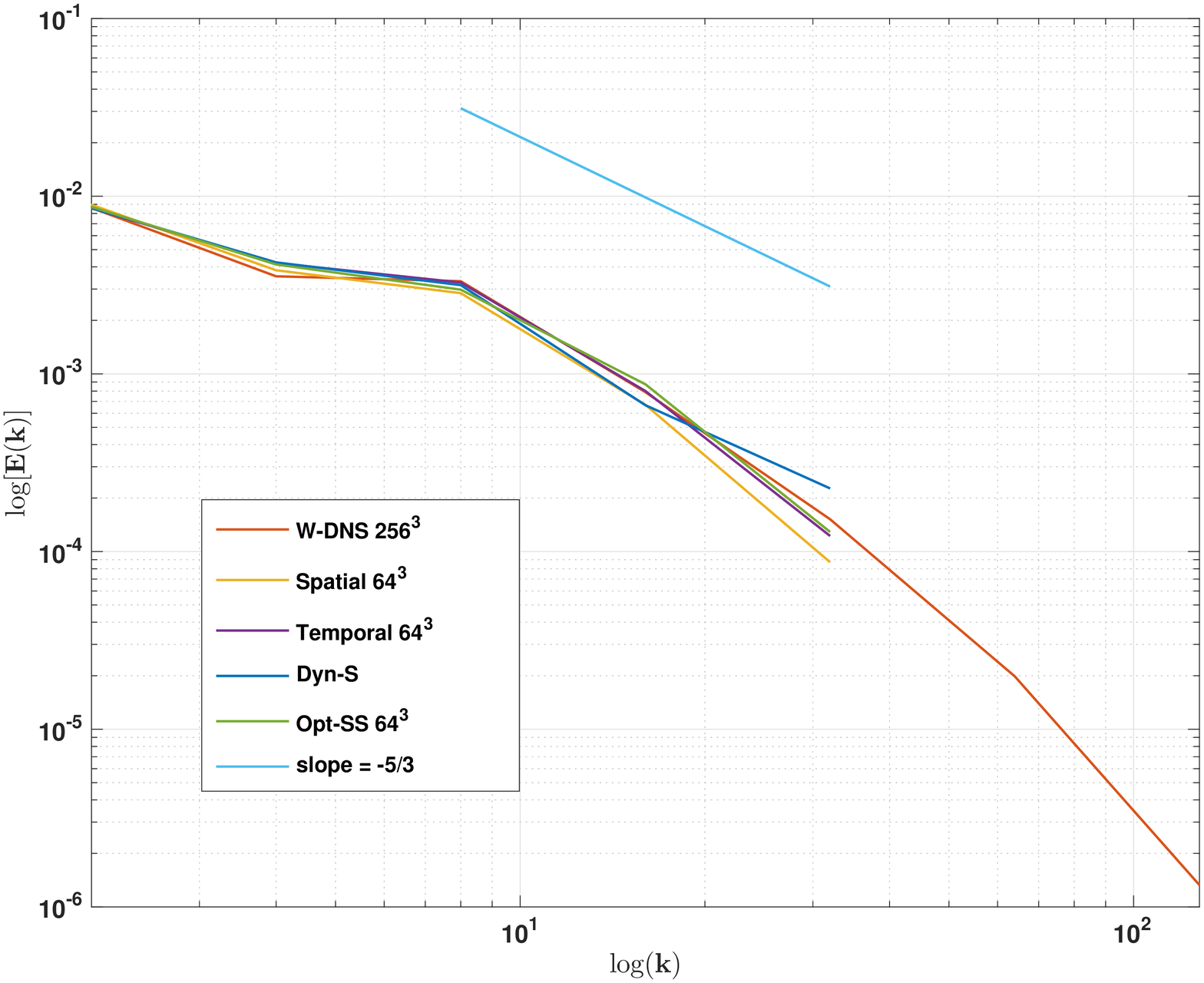}\includegraphics[width=0.55\textwidth]{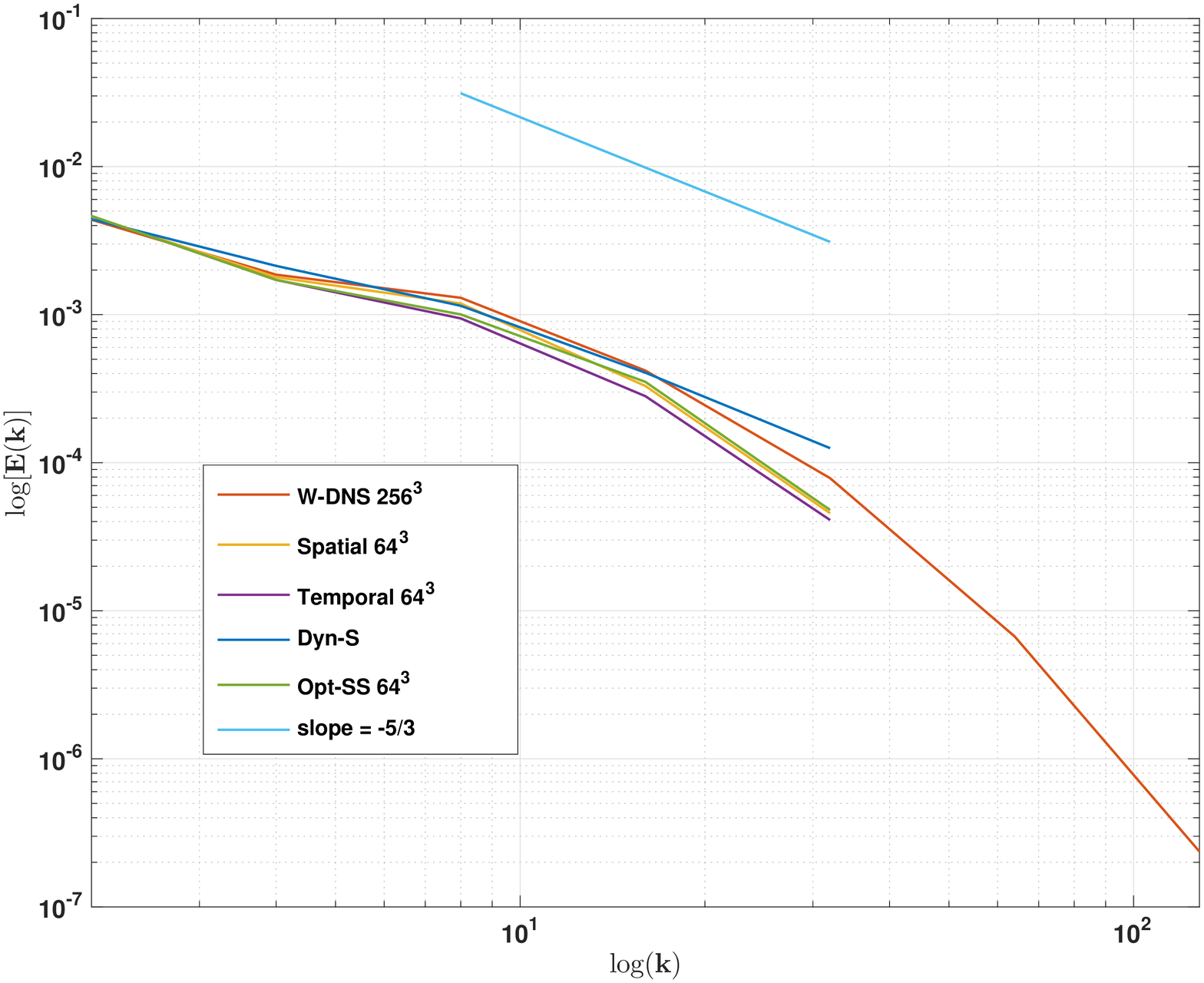}
\caption{Wavelet energy spectrum at $t\approx9.61$ (left) and $t\approx 14.7$ (right). Comparison between the three variance tensor  models, the dynamic Smagorinsky model and the DNS solution. Taylor-Green vortex at $Re=1600$.}
\label{WSpectreLES2}
\end{figure}

As we rely on a wavelet  scheme for the numerical simulations, it is insightful to inspect the discrete power spectra computed from the wavelet coefficients. They are plotted on figures \ref{WSpectreLES1} and \ref{WSpectreLES2}.  Wavelet power spectrum corresponds to an averaged version of the Fourier spectrum \citep{Bruneau02} and exhibits the same slopes as the Fourier spectrum \citep{Perrier95}. A discrete version of the wavelet spectrum as  plotted here provides one energy measure per scale level. It can be  observed on the four spectra that the dynamic Smagorinsky tensor exhibits an undue energy intake at the cutoff scale. This  amplification of energy is likely due to noisy velocity fields at the cutoff. The three different model ensuing from our statistical representation of the small-scale velocity component does not show such a deficiency.  


\begin{figure}
\centering
\hspace*{-1cm}
\includegraphics[width=0.55\textwidth]{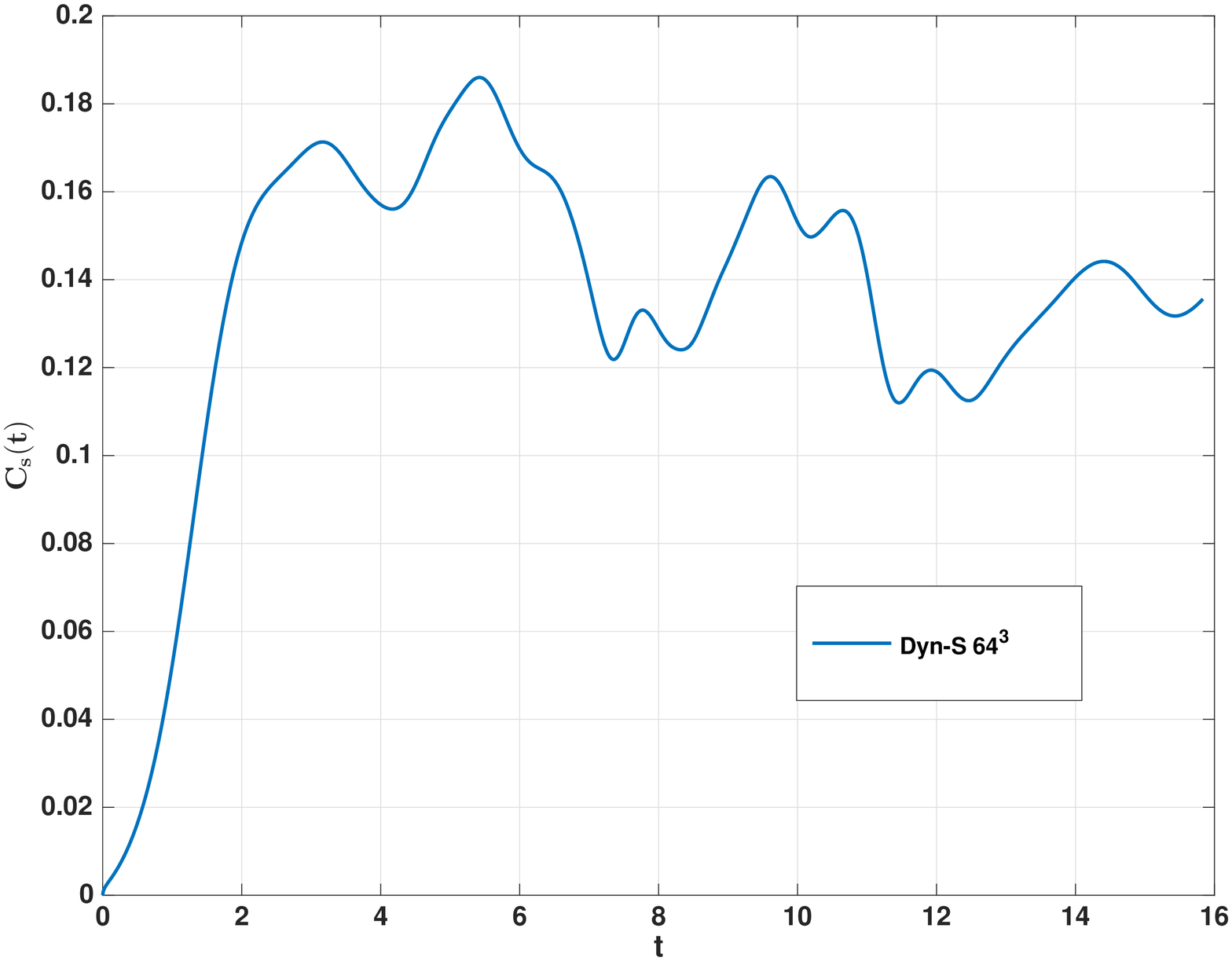}
\caption{Time evolution of the dynamic constants of the Smagorinsky model. Taylor-Green vortex simulation at $Re=1600$ and $L=6$.}
\label{SmgCtFig1600}
\end{figure}

For information purpose we draw on figure \ref{SmgCtFig1600} the value along time of the constant weighting the dynamic Smagorinsky model. The obtained maximum value is about $0.1860$ and the mean value is $0.1366$, this is in good agreement with the predicted values, see \citep{Germano91}. 

\

Regarding the different criteria explored,  we see that the proposed  large-scale subgrid tensors based on local velocity covariances  perform better than the dynamic Smagorinsky  subgrid  model. Among all the models proposed, the optimal variance tensor estimation based on a two-scale similarity assumption and the variance tensor constructed from a local temporal variance of the resolved velocity field outperform significantly the others. They lead to very good results that are strikingly close to each other. To our knowledge they both outperform the state-of-the-art solutions for this flow \citep[and references therein]{Chapelier13,Rees11}. Though very simple,  the temporal covariance model weighted by a unique constant fixed through the quadratic scaling rule presented section \ref{Section-Scaling}  gives very good results. This demonstrates the pertinence of this scaling. The choices operated here are among the simplest that can be devised. More involved schemes could be easily imagined. For instance, variance tensors based on vorticity statistics might be very interesting to explore. Another route would be to elaborate this tensor from statistics  extracted from measurements or DNS data. Large-eddies simulation models could likely be proposed in this spirit for non-homogeneous turbulence or boundary layer flows.  Note also that both correlation schemes rely on a constant fixed through a rough dimensional scaling. A dynamical procedure in the same vein as the one used for the dynamics Smagorinsky subgrid model could be beneficial to get a finer estimate of the constant. As our  purpose was here to bring a simple 
demonstration of the wide potentiality offered by the proposed stochastic modeling, we leave such  potential improvements to future works.  

\section{Conclusion}

In this paper we have described a decomposition of the Navier-Stokes equations in terms of a temporally smooth  velocity component and a fast oscillating random field associated to the unresolved flow component. This decomposition leads to a new large-scale representation paradigm, which can be interpreted as a large eddies simulation formalized through a time-scale separation.  An advection correction and a subgrid diffusion term both emerge from this formalism. They encode respectively {\em turbophoresis} phenomenon  and anisotropic mixing effect due to turbulence. The corresponding subgrid tensor enables generalizing  the Boussinesq eddy viscosity concept on solid a theoretical ground. Such large-scale representation have been assessed on a Taylor-Green vortex flow. We compared different models of the variance tensor built from local averaging and  a scale similarity least squares estimation  procedure. The different numerical simulations outperform a standard dynamic Smagorinsky model based on Boussinesq's eddy viscosity assumption. We believe those first results constitute are very encouraging as more advanced models could yet yield substantial improvements. As a matter of fact this formalism, built from physical conservation laws, paves the way to new possibilities to design efficient subgrid schemes. One could for instance explore models where  the diffusion tensor is learned from DNS data or from small-scale observations. Another route of investigation consists in defining adapted basis for the small-scale random field from the fluctuations observed on two consecutive scales of the resolved tensor. 

%
 
\bibliographystyle{plain}

\begin{thebibliography}{10}


\bibitem{Aspden08}
A.~Aspden, N.~Nikiforakis, S.~Dalziel, and J.B. Bell.
\newblock Analysis of implicit les methods.
\newblock {\em App. Math. Comp. Sci.}, 3(1):103--126, 2008.

\bibitem{Bardina80}
J.~Bardina, J.H. Ferziger, and W.C. Reynolds.
\newblock Improved subgrid scale models for large eddy simulation.
\newblock volume~80, page 1357.

\bibitem{Bensoussan-Temam-73}
A.~Bensoussan and R.~Temam.
\newblock Equations stochastique du type {N}avier-{S}tokes.
\newblock {\em J. Funct. Anal.}, 13, 1973.

\bibitem{Boris92}
J.P. Boris, F.F. Grinstein, E.S. Oran, and R.L. Kolbe.
\newblock New insights into large-eddy simulation.
\newblock {\em Fluid Dynamics Research}, 10:199--228, 1992.

\bibitem{Boussinesq77}
J.~Boussinesq.
\newblock Essai sur la th\'eorie des eaux courantes.
\newblock M\'emoires pr\'esent\'es par divers savants \`a l'Acad\'emie des
  Sciences, 23 (1): 1--680, 1877.

\bibitem{Brachet83}
M.~Brachet, D.~Meiron, S.~Orszag, B.~Nickel, R.~Morf, and U.Frisch.
\newblock Small-scale structure of the taylor-green vortex.
\newblock {\em J. Fluid. Mech.}, 130(6):411--452, 1983.

\bibitem{Brooke92}
M.~Brooke, K.~Kontomaris, T.~Hanratty, and J.~McLaughlin.
\newblock Turbulent deposition and trapping of aerosols at a wall
\newblock {\em Phys. of Fluids}, A 4, 825–834,1992.


\bibitem{Bruneau02}
Ch-H.~ Bruneau, P.~Fischer, Z.~Peter, and  A.~Yger
\newblock Comparison of numerical methods for the computation of energy spectra in 2D turbulence. Part I: Direct methods
\newblock {\em Sampling Theory Sign. and Im. Proc.}, 1(1):0:50, 2002.


\bibitem{Buizza99}
R.~Buizza, M.~Miller, and T.N. Palmer.
\newblock Stochastic representation of model uncertainties in the ECMWF
  ensemble prediction system.
\newblock {\em Quarterly Journal Royal Meteorological Society}, 125:2887--2908,
  1999.

\bibitem{Chacon-Lewandowski}
T.~Chacon-Rebollo and R.~Lewandowski.
\newblock {\em Mathematical and numerical foundations of turbulence models and
  applications}.
\newblock Springer New-York, 2014.

\bibitem{Chapelier13}
J.B. Chapelier, M.~De La~Llave Plata, F.~Renac, and E.~Martin.
\newblock Final abstract for {ONERA} {T}aylor- {G}reen {DG} participation.
\newblock In {\em 1st International Workshop on High-Order CFD Methods at the
  50th AIAA Aerospace Sciences Meeting}, Nashville,TN, 2013.

\bibitem{Caporaloni75}
M. Caporaloni, F. Tampieri, F. Trombetti, and O. Vittori
\newblock Transfer of particles in non- isotropic air turbulence.
\newblock {\em J. Atmos. Sci}, 32:565–568,1975.

\bibitem{Dairay16}
T.~Dairay, E.~Lamballais, S.~Laizet, and J.C. Vassilicos.
\newblock Numerical dissipation vs. subgrid-scale modeling for large eddy
  simulation.
\newblock to be published, 2016.

\bibitem{Deardorff70}
J.~Deardorff.
\newblock A numerical study of three-dimensional turbulent channel flow at
  large reynolds numbers.
\newblock {\em J. Fluid Mech.}, 1970.

\bibitem{Deriaz06}
E.~Deriaz and V.~Perrier.
\newblock Divergence-free and curl-free wavelets in 2d and 3d, application to
  turbulent flows.
\newblock {\em J. of Turbulence}, 7(3):1--37, 2006.

\bibitem{Franzke15}
C.~Franzke, T.~O'Kane, J.~Berner, P.~Williams, and V.~Lucarini.
\newblock Stochastic climate theory and modeling.
\newblock {\em Wiley Interdisciplinary Reviews: Climate Change}, 6(1):63--78,
  2015.
  

\bibitem{Gent90}
P.~Gent and J.~McWilliams.
\newblock Isopycnal mixing in ocean circulation models.
\newblock {\em J. Phys. Oceanogr.}, 20:150--155, 1990.


\bibitem{Germano92}
M.~Germano.
\newblock Turbulence : the filtering approach.
\newblock {\em J. Fluid Mech.}, 1992.

\bibitem{Germano91}
M.~Germano, U.~Piomelli, P.~Moin, and W.~H. Cabot.
\newblock A dynamic subgrid-scale eddy viscosity model.
\newblock {\em Phys. of Fluids}, 3:1760--1765, 1991.

\bibitem{Ghosal96}
S.~Ghosal.
\newblock An analysis of numerical errors in large-eddy simulations of
  turbulence.
\newblock {\em J. Comp. Phys.}, 125:187--206, 1996.

\bibitem{Ghosal95}
S.~Ghosal and P.~Moin.
\newblock The basic equations for the large eddy simulation of turbulent flows
  in complex geometry.
\newblock {\em J. Comp. Phys.}, 1995.

\bibitem{Gronskis13}
A.~Gronskis, D.~Heitz, and E.~M{\'e}min.
\newblock Inflow and initial conditions for direct numerical simulation based
  on adjoint data assimilation.
\newblock {\em J. Comp. Phys}, 242(6):480--497, 2013.

\bibitem{Grooms13}
I.~Grooms and A.~Majda.
\newblock Efficient stochastic superparameterization for geophysical
  turbulence.
\newblock {\em PNAS}, 110(12), 2013.

\bibitem{Haworth86}
D.~Haworth and S.~Pope.
\newblock A generalized Langevin model for turbulent flows.
\newblock {\em Phys. of Fluids}, 29:387--405, 1986.

\bibitem{Laval06}
B.~Dubrulle J.-P~Laval and J.C. McWilliams.
\newblock Langevin models of turbulence: Renormalization group, distant
  interaction algorithms or rapid distortion theory?
\newblock {\em Pys. of Fluids}, 15(5):1327--1339, 2006.

\bibitem{Kadri-IJCV-13}
S.~Kadri-Harouna, P.~D\'erian, P.~H\'eas, and E.~M\'emin.
\newblock Divergence-free wavelets and high order regularization.
\newblock {\em International Journal of Computer Vision}, 103(1):80--99, 2013.

\bibitem{Kadri13}
S.~Kadri-Harouna and V.~Perrier.
\newblock Effective construction of divergence-free wavelets on the square.
\newblock {\em J. of Computational and Applied Math.}, 2013.

\bibitem{Karamanos00}
G.~Karamanos and G.~Karniadakis.
\newblock A spectral vanishing viscosity method for large-eddy simulations.
\newblock {\em Journal of Computational Physics}, 163(1):22--50, 2000.

\bibitem{Kraichnan59}
R.~Kraichnan.
\newblock The structure of isotropic turbulence at very high reynolds numbers.
\newblock {\em J. of Fluids Mech.}, 5:477--543, 1959.

\bibitem{Kraichnan68}
R.~Kraichnan.
\newblock Small-scale structure of a scalar field convected by turbulence.
\newblock {\em Phys. of Fluids}, 11:945--963, 1968.

\bibitem{Kraichnan70}
R.~Kraichnan.
\newblock Convergents to turbulence functions.
\newblock {\em J. of Fluid Mech.}, 41:189--217, 1970.

\bibitem{Kunita}
H.~Kunita.
\newblock {\em Stochastic flows and stochastic differential equations}.
\newblock Cambridge University Press, 1990.

\bibitem{Lamballais11}
E.~Lamballais, V.~Fortun\`e, and S.~Laizet.
\newblock Straightforward high-order numerical dissipation via the viscous term
  for direct and large eddy simulation.
\newblock {\em J. Comp. Phys.}, 230:3270--3275, 2011.

\bibitem{Layton00}
W.~Layton.
\newblock Approximating the larger eddies in fluid motion v: Kinetic energy
  balance of scale similarity models.
\newblock {\em Math. and Comp. Modelling}, 2000.

\bibitem{Leith71}
C.~Leith.
\newblock Atmospheric predictability and two-dimensional turbulence.
\newblock {\em J. Atmos. Sci}, 28, 1971.

\bibitem{Leith90}
C.~Leith.
\newblock Stochastic backscatter in a subgrid-scale model: plane shear mixing
  layer.
\newblock {\em Phys. of Fluids}, 2(3):1521--1530, 1990.

\bibitem{Lilly66}
D.~Lilly.
\newblock On the application of the eddy viscosity concept in the inertial
  subrange of turbulence.
\newblock Technical Report 123, NCAR, 1966.

\bibitem{Lilly92}
D.~Lilly.
\newblock A proposed modification of the {G}ermano subgrid-scale closure.
\newblock {\em Phys. Fluids}, 3:2746--2757, 1992.

\bibitem{Macinnes92}
J. Macinnes and F. Bracco.
\newblock Stochastic particles dispersion modelling and the tracer-particle limit.
\newblock {\em Phys. Fluids}, 4(12):2809--2824, 1992.

\bibitem{Majda99}
A.~Majda, I.~Timofeyev, and E.~Vanden Eijnden.
\newblock Models for stochastic climate prediction.
\newblock {\em PNAS}, 1999.

\bibitem{Majda03}
A.~Majda, I.~Timofeyev, and E.~Vanden Eijnden.
\newblock A systematic strategies for stochastic mode reduction in climate.
\newblock {\em Journ. Atmos. Sci.}, 60:1705--1722, 2003.

\bibitem{Mason92}
P.J. Mason and D.J. Thomson.
\newblock Stochastic backscatter in large-eddy simulations of boundary layers.
\newblock {\em J. of Fluid Mech.}, 242:51--78, 1992.

\bibitem{Memin14}
E.~M{\'e}min.
\newblock Fluid flow dynamics under location uncertainty.
\newblock {\em Geophysical \& Astrophysical Fluid Dynamics}, 108(2):119--146,
  2014.

\bibitem{Meneveau00}
C.~Meneveau and J.~Katz.
\newblock Scale-invariance and turbulence models for large-eddy simulation.
\newblock {\em Annu. Rev. Fluid. Mech}, 32:1--32, 2000.


\bibitem{Minier14}
J.-P.~Minier S.~Chibbaro and S.~Pope.
\newblock Guidelines for the formulation of Lagrangian stochastic models for particle simulations of single-phase and dispersed two-phase turbulent flows.
\newblock {\em Phys. of Fluids}, 26,1113303, 2014.


\bibitem{Monin-YaglomB}
A.S. Monin and A.M. Yaglom.
\newblock {\em Statistical fluid mechanics}, volume~II.
\newblock MIT Press, 1975.

\bibitem{Nocedal99}
J.~Nocedal and S.J. Wright.
\newblock {\em Numerical optimization}.
\newblock Springer Series in Operations Research. Sringer-Verlag, New-York,
  1999.

\bibitem{Orszag70}
S.~Orszag.
\newblock Analytical theories of turbulence.
\newblock {\em J. Fluid Mech.}, 41:363--386, 1970.

\bibitem{Orszag74}
S.~Orszag.
\newblock Numerical simulation of the taylor-green vortex.
\newblock In Springer-Verlag, editor, {\em International Symposium on Computing
  Methods in Applied Sciences and Engineering}, volume~2, pages 50--64, 1974.

\bibitem{Palmer08}
T.~Palmer and P.~Williams.
\newblock Theme issue 'stochastic physics and climate modelling'.
\newblock {\em Phil. Trans. R. Soc.}, 366(1875), 2008.

\bibitem{Pasquetti06}
R.~Pasquetti.
\newblock Spectral vanishing viscosity method for large-eddy simulation of
  turbulent flows.
\newblock {\em J. Sci. Comp.}, 27(1-3):365--375, 2006.

\bibitem{Perrier95}
V~Perrier, T.~Philipovitch, and C.~Basdevant,.
\newblock Wavelet spectra compared to Fourier spectra
\newblock {\em J. Math. Phys.}, 36:1506–1519, 1995.

\bibitem{Piomelli91}
U.~Piomelli, W.~Cabot, P.~Moin, and S.~Lee.
\newblock Subgrid-scale backscatter in turbulent and transitional flows.
\newblock {\em Phys. Fluids}, 3(7):1766--1771, 1991.

\bibitem{Pope94}
S.~Pope.
\newblock Lagrangian PDF methods for turbulent flows.
\newblock {\em Annu. Rev. Fluid Mech.}, 26:23-63, 1994.

\bibitem{Pope00}
S.~Pope.
\newblock {\em Turbulent flows}.
\newblock Cambridge University Press, 2000.

\bibitem{Prandtl25}
L.~Prandtl.
\newblock Bericht uber untersuchungen zur ausgebildeten turbulenz.
\newblock {\em Angew. Math, Meth.}, 5:136--139, 1925.

\bibitem{DaPrato}
G.~Da Prato and J.~Zabczyk.
\newblock {\em Stochastic equations in infinite dimensions}.
\newblock Cambridge University Press, 1992.

%

\bibitem{Resseguier16a}
V.~Resseguier, E.~M{\'e}min, and B.~Chapron.
\newblock Geophysical flows under location uncertainty, Part I Random transport and general models
\newblock  \url{https://hal.inria.fr/hal-01391420}, 2016.

\bibitem{Resseguier16b}
V.~Resseguier, E.~M{\'e}min, and B.~Chapron.
\newblock Geophysical flows under location uncertainty, Part II Quasi-geostrophy and efficient ensemble spreading
\newblock \url{ https://hal.inria.fr/hal-01391476}, 2016.

\bibitem{Resseguier16c}
V.~Resseguier, E.~M{\'e}min, and B.~Chapron.
\newblock Geophysical flows under location uncertainty, Part III SQG and frontal dynamics under strong turbulence conditions
\newblock \url{https://hal.inria.fr/hal-01391484}, 2016.

\bibitem{Sagaut05}
P.~Sagaut.
\newblock {\em Large-eddy simulation for incompressible flow - An introduction,
  third edition}.
\newblock Springer-Verlag, Scientic Computation series, 2005.


\bibitem{Sawford86}
B.~Sawford.
\newblock Generalized random forcing in random-walk models of turbulent dispersion models.
\newblock {\em Phys. of Fluids.}, 29:3582--3585, 1986.

\bibitem{Schumann95}
U.~Schumann.
\newblock Stochastic backscatter of turbulence energy and scalar variance by
  random subgrid-scale fluxes.
\newblock {\em Proc. R. Soc. Lond. A}, 451:293--318, 1995.

\bibitem{Sehmel70}
G.~Sehmel.
\newblock Particle deposition from turbulent air flow.
\newblock {\em J. Geophys. Res.}, 75:1766--1781, 1970.

\bibitem{Shutts05}
G.~Shutts.
\newblock A kinetic energy backscatter algorithm for use in ensemble prediction
  systems.
\newblock {\em Quarterly Journal of the Royal Meteorological Society},
  612:3079--3012, 2005.

\bibitem{Smagorinsky63}
J.~Smagorinsky.
\newblock General circulation experiments with the primitive equation: I. the
  basic experiment.
\newblock {\em Monthly Weather Review}, 91:99--165, 1963.

\bibitem{Tadmor89}
E.~Tadmor.
\newblock Convergence of spectral methods for nonlinear conservation laws.
\newblock {\em SIAM J. Numer. Anal}, 26(1):30--44, 1989.

\bibitem{Taylor-Green}
G.~Taylor and A.~Green.
\newblock Mechanisms of the production of small eddies from large ones.
\newblock {\em Proceedings of the Royal Society of London A}, 1938.

\bibitem{Rees11}
W.M. Van~Rees, D.I. Leonard, A.and~Pullin, and P.~Koumoutsakos.
\newblock A comparison of vortex and pseudo-spectral methods for the simulation
  of periodic vortical flows at high reynolds numbers.
\newblock {\em J. Comp. Phys}, 230:2794--2805, 2011.


\end{thebibliography}

%
%
%
%
%
%
%
%
%
%
\clearpage
\appendix
\section{Stochastic Reynolds transport theorem}
\label{appendix:srt}
We derive in this appendix  the expression of the Reynolds transport theorem for the drift plus noise decomposition:
In a Lagrangian stochastic picture, the infinitesimal displacement associated to a trajectory $X_t$ of a particle  is noted:
\begin{equation}
\label{particle_dX}
\dif\XX_t = \w(\XX_t,t)dt + \mbs \sigma(\XX_t,t) \dif\B_t.
\end{equation}
The random field involved in this equation  is defined  for all points of the fluid domain, $\Omega$ through the kernel $\breve{\mbs \sigma} (.,.,t) $ associated to the diffusion operator $\mbs \sigma(.,t)$
 \begin{equation}
(\mbs \sigma(\xx,t)\mbs f)^i \defin \sum_{j=1}^d\int_\Omega \breve{\sigma}^{ij}(\xx,\yy,t)  f^j (\yy,t) \dif\yy.
\end{equation}
This operator is assumed to be  of finite norm for any orthonormal basis of the associated Hilbert space ${\cal H}^d$) and to have a null boundary condition on the domain frontier: 
\begin{equation}
\mbs \sigma(\xx,t) \mbs f =0 \mbox{ on } \partial\Omega,\;  \forall \mbs f \in {\cal H}^d.
\end{equation}
Function $\B_t$ denotes a $d$-dimentional Brownian function (see \cite{DaPrato}  for more information on infinite dimensional Wiener process) and the resulting $d$-dimensional random field, $\mbs \sigma(\xx,t) \dif \B_t \in {\cal H}^d$, is a centered vectorial Gaussian function correlated in space and uncorrelated in time with covariance tensor:
\[
 Q_{ij}(\xx,\yy,t,t')= \int_\Omega \breve{\mbs \sigma}^{ik}(\xx,\yy',t) \;\breve{\mbs \sigma}^{jk}(\yy',\yy,t)\dif\yy' \delta(t-t')\dif t.
\]
We observe that those random fields have a (mean) bounded norm: $\Exp \|\mbs \sigma d\B_t\|^2= tr \;\mbs Q <\infty$, where the trace of the covariance tensor is given for any basis $\{{\mbs e}_k \;|\;k\in \N\}$ of ${\cal H}^d$ as $tr\;{\mbs Q}= \sum_k ({\mbs e}_k,  \mbs Q {\mbs e}_k)_{\cal H}$. In the following we will note the diagonal of the covariance tensor as: $\mbs a (\xx) \dif t= \mbs Q(\xx,\xx)$. Tensor, $\mbs a$, that will be referred to as the variance tensor
\begin{equation*}
 \mbs a(\xx,t)\delta(t-t')\dif t= \Exp \left ( \left (\mbs \sigma(\xx,t) \dif\B_t \right ) \left ( \mbs \sigma(\xx,t') \dif\B_{t'} \right ) \transp \right ),
\end{equation*}
   is by definition a symmetric positive definite matrix at all spacial points, $\xx$. This quantity corresponds to the time derivative of the so-called quadratic variation process:
\begin{align*}
\mbs a(\xx,t) \dif t  
&= \int_{\Omega} \breve{\mbs \sigma} (\xx,\zz) \breve{\mbs \sigma}\transp  (\xx,\zz) \dif \zz dt
\defin \mbs \sigma(x) \mbs \sigma(y)\transp \dif t \\
&= \dif \left < \int_0^t  \mbs \sigma(\xx,s) \dif \B_{s}, \int_0^t  \mbs \sigma(\xx,r) \dif \B_{r}\right >.
\end{align*}
The notation $\left<f,g\right>$ stands for  the quadratic cross-variation process of $f$ and $g$. This central object of Stochastic Calculus, can be interpreted as the covariance along time of the increments of $f$ and $g$. The quadratic variation process is briefly presented in Appendix \ref{QuadVar}. 

The drift term, $\w$,  of  Lagrangian expression (\ref{particle_dX}), represents the "smooth" -- differentiable -- part of the flow whereas the random part, 
\begin{equation}
\dot{\mbs W}=\mbs \sigma \frac{\dif\B_t}{\dif t},
\end{equation} figures the small-scale velocity component associated to a much thinner time-scale. This component  is modeled as a delta-correlated  random field in time as it represents a highly non regular process at the resolved time scale. In  this work, we assume that this small-scale random component is incompressible and therefore associated to a divergence free diffusion tensor: $\nab\bcdot \mbs \sigma=0$. This assumption, which  obviously does not prevent the drift component (and therefore the whole field) to be compressible, leads to much simpler modeling and remains realistic for the models considered in this study.

Let us consider now a  spatially  regular enough scalar function $\phi$ of compact support, transported by the stochastic flow (\ref{particle_dX}) and that vanishes outside volume ${\cal V}(t)$ and on its boundary $\partial{\cal V}(t)$. As this function is assumed to be transported by the stochastic flow, it constitutes a stochastic process defined from its initial time value  $g$: 
\[
\phi(\XX_t,t)= g(\xx_0).
\]
We will assume that both functions have bounded spatial gradients. Besides, the initial function $g:\Omega\rightarrow \reel$ vanishes outside the initial volume ${\cal V}(t_0)$ and on  its boundary.  
Let us point out that in this construction, function $\phi$ cannot be a deterministic function. As a matter of fact, if it was the case, its differential would be given by a standard Ito formula
\begin{equation}
 \biggl(\partial_t\phi +{\nab}\phi\, {\bm\cdot}\, \w +  \frac{1}{2} \sum_{i,j=1}^d \dif\bigl\langle X^{i}_t,X^{j}_t\bigr\rangle \frac{\partial^2 \phi }{\partial x_i \partial x_j}\biggr) \dif t+ {\nab}\phi \,{\bm\cdot}\, \mbs \sigma \dif \hat\B_t =0,
\end{equation}
which here cancels as $\phi$ is transported by the flow. A separation between the slow deterministic terms and the fast Brownian term would yield  a specific noise term orthogonal to $\nab \phi$. This would boil down to the deterministic case, which is not the general goal followed here. 

As $\phi$ is a random function, the differential of  $\phi(\XX_t,t)$  involves the composition of two stochastic processes. Its evaluation requires the use of a generalized Ito formula  usually referred in the literature to as the Ito-Wentzell formula \citep[see theorem 3.3.1,][]{Kunita}.
This extended Ito formula incorporates in the same way as the classical Ito formula a quadratic variation terms related to process $\XX_t$ but also co-variation terms between $\XX_t$ and the gradient of the random function $\phi_t$. Its expression is in our case given by
\begin{align}
\dif\phi(t,\XX_t)= &\,\dif_t\phi +{\nab}\phi \,{\bm\cdot}\, \dif\XX_t +\sum_i \dif\Bigl\langle\frac{\partial \phi}{\partial x_i}, X^i_t\Bigr\rangle \dif t \nonumber+ \frac{1}{2} \sum_{i,j} \dif\bigl\langle X^{i}_t,X^{j}_t\bigr\rangle \frac{\partial^2 \phi }{\partial x_i \partial x_j} \dif t \nonumber\\= &\,0.
\label{dxi=0}
\end{align}
 It can be immediately checked that for a deterministic function, the standard Ito formula is recovered since the co-variations terms between $\XX_t$ and $\nab \phi_t$ cancel. 
 
 It follows from \ref{dxi=0} that for a fixed grid point, function $\phi(\xx,t)$ is solution of a stochastic differential equation of the form
\begin{equation}
\dif_t\phi(\xx,t) = v(\mbs x,t)\dif t + \mbs f(\mbs x,t)\,{\bm\cdot}\, \dif \B_t,
\label{dxi_t}
\end{equation}
where the Brownian term must compensate the Brownian term of (\ref{dxi=0}).
The quadratic variation term involved in (\ref{dxi=0}) is given through (appendix \ref{QuadVar}) as
\begin{align}
\dif\bigl\langle X^{i}_t,X^{j}_t\bigr\rangle  &= a^{ij}(\xx,t)  \\\nonumber
&= \sum_{k}  \mbs \sigma^{ik}(\xx,t)\mbs \sigma^{kj}(\xx,t)\\&= \sum_{k} \int_\Omega \breve \sigma^{ik}(\xx,\yy,t)   \breve\sigma^{kj} (\yy,\xx,t),
 \end{align} and the covariation term reads
\begin{align}
\dif\Bigl\langle\frac{\partial \phi_t}{\partial x_i}, X_t^i\Bigr\rangle = \sum_{j} \int_\Omega \frac{\partial \breve f^j}{\partial x_i}(\xx,\yy,t)   \breve\sigma^{ij} (\yy,\xx,t).
\end{align}
In these expressions $\breve {\mbs f}$ (resp.  $\breve{\mbs \sigma}$) designates  the kernel associated to operator $\mbs f$ (resp.  $\mbs \sigma$).
Now, identifying first the Brownian term and next the deterministic terms  of  equations (\ref{dxi=0}) and (\ref{dxi_t}), we infer 
\begin{align}
\breve{\mbs  f}(\xx,\yy,t)\transp & = - {\nab}\phi(\xx,t) \transp \breve{\mbs \sigma}(\xx,\yy,t),\nonumber\\
v(\xx,t) &= - {\nab}\phi\,{\bm\cdot}\, \w + \sum_{i,j}\frac{1}{2} a_{ij}\frac{\partial^2 \phi}{\partial x_i \partial x_j} + \nab \phi\,{\bm\cdot}\, \frac{\partial \mbs \sigma^{\bullet j}} {\partial x_i} \mbs \sigma^{ij}, \nonumber\\
&= - {\nab}\phi\,{\bm\cdot}\, \w +  \nab \phi \, {\bm\cdot} \,(\nab{\bm\cdot} \,\mbs a)\transp +   \frac{1}{2}\sum_{i,j} a_{ij}\frac{\partial^2 \phi}{\partial x_i \partial x_j}, \nonumber
\end{align}
and finally get
\begin{align}
\label{SMD}
\dif_t\phi   &=  {\cal L} \phi  \dif t  -  {\nab}\phi\,{\bm\cdot}\,\mbs \sigma \dif\B_t, \\
{\cal L} \phi &= -{\nab}\phi\,{\bm\cdot}\,\bigl(\w - \frac{1}{2}\,(\nab{\bm\cdot} \,\mbs a)\transp\bigr)  + \frac{1}{2} \nab{\bm\cdot} \, (\mbs a \nab \phi).\label{defL}
\end{align}
This differential at a fixed point, $\xx$, defines the equivalent in the deterministic case of the material derivative of a function transported by the flow. 
The differential of the integral over a material volume of the product $q\phi$ is given by
\begin{align}
\dif\int_{{\cal V}(t)} \!\!\!\!q\phi(\XX_t,t)\dif \xx&= \dif\int_{\Omega} \!\!q\phi \dif \xx,\nonumber\\
&= \int_{\Omega} \!\bigl(\dif_t q \phi + q \dif_t\phi  +\dif_t\langle q,\phi\rangle \bigr) \dif \xx, \nonumber
\end{align}
where the first line comes from $\phi(t,\xx)= 0 \mbox{ if }\xx\in\Omega\backslash {\cal V}(t)$
and the second one from the integration by part formula of two  Ito processes. 
Hence, from (\ref{SMD}) this differential is 
\begin{equation*}
 \int_{\Omega} \biggl(\dif_t q \phi + \bigl(q {\cal L} \phi + \nab \phi\, {\bm\cdot}\, \mbf a \nab q \bigr)\dif t\, \dif \xx- \int_{\Omega} q {\nab}\phi\,{\bm\cdot} \, \mbs \sigma \dif \B_t.
\end{equation*}
Introducing ${\cal L}^\ast$ the (formal) adjoint  of the operator ${\cal L}$ in the space $L^2(\Omega)$ with Dirichlet boundary conditions, this expression can be written as
\begin{equation}
\int_{\Omega} \!\!\left( \dif_t q + \bigl({\cal L}^\ast q -\nab{\bm\cdot} \,(\mbf a \nab q )\bigr)\dif t + \!\nab{\bm\cdot}\bigl(q \mbs \sigma \dif\B_t\bigr) \right)\phi\, \dif \xx.\nonumber
\end{equation}
With the complete expression of ${\cal L}^\ast$ (remarking that the second right-hand term of \ref{defL} is self-adjoint) and  if $\phi(\xx,t) \rightarrow \car_{{\cal V}(t) /\partial{\cal V}(t)}$, where $\car$ stands for the characteristic  function, we get the sought form of this differential:
\begin{equation}
\int_{{\cal V}(t)}\! \biggl[\dif_t q +\biggl( {\nab} {\bm\cdot}\,\bigl( q\w - \frac{1}{2}\,\nab{\bm\cdot} \,(q\mbs a)\transp\bigr)  \biggr) \dif t + \nab q \mbs \sigma \dif\B_t\bigr)\biggr]\dif \xx,\nonumber
\end{equation}

\section{Quadratic variation and covariation}
\label{QuadVar}
We recall here the notion of quadratic variation and co-variation, which play a central role in stochastic calculus. We will here restrict ourselves to standard Ito processes.   Quadratic variation and co-variation correspond respectively to the variance and covariance   of the process  increments along time. The quadratic co-variation process denoted as $\langle\XX,\YY\rangle _t$, (respectively the quadratic variation for $\YY=\XX$) is defined as the limit in probability over a partition $\{t_1,\ldots,t_n\}$ of $[0,t]$ with $t_1<t_2<\cdots<t_n$, and a partition spacing $\delta t_i= t_i -t_{i-1}$, noted as $|\delta t|_n= \max\limits_{i} \delta t_i$ and such that $|\delta t|_n\to 0$ when $n\to\infty$:

\begin{equation}
\nonumber
\langle\XX,\YY\rangle _t = \lim^P_{|\delta t|_n\rightarrow 0} \sum_{i=0}^{n-1} \bigl(\XX(t_{i+1}) - \XX(t_i)\bigr) \bigl(\YY(t_{i+1}) - \YY(t_i)\bigr)\transp.
\end{equation}
For Brownian motion these covariations can be easily computed and are given by the following rules $\langle B,B\rangle _t =t$, $\langle B,h\rangle _t=\langle h,B\rangle _t = \langle h,h\rangle _t=0$, where $h$ is a deterministic function and $B$ a scalar Brownian motion.

\section{General properties of the large-scale stochastic Navier-Stokes model}\label{Prop-NSU}
First of all, it is important to emphasize  that the distribution of the  velocity anomaly, $\mbs U(\xx,t) -\Exp \mbs U(\xx,t)$,  with the Eulerian velocity field 
\begin{equation}
{\mbs U} (\xx,t) = \w(\xx,t)  + \dot{\mbs W}(\xx,t)
\end{equation}
 is, in the general case, {\em  not} Gaussian. As a matter of fact, due to the multiplicative noise (the diffusion tensor depends {\em a priori} on the flow) and as the resolved dynamics is  nonlinear, the ensemble mean of the flow velocity is in general not given by  $\w(\xx,t)$. Hence, this construction does not correspond to a  turbulence model with a Gaussian closure hypothesis of the fourth-order correlation functions as considered in the Millionschikov hypothesis \citep{Monin-YaglomB}. Nor is it based on quasi-normal approximations such as the EDQNM approaches \citep{Orszag70}, which opposite to the previous hypothesis ensures the positivity of  the energy spectrum. In the proposed decomposition, the velocity is clearly a Markovian stochastic process, but its distribution is not Gaussian.  Note also that in our approach no isotropic assumption on the increments nor on the random component are considered to define the subgrid tensor.
  
 As already mentioned, the proposed stochastic representation can be interpreted as a temporal decomposition of the original Navier-Stokes equations. In the following sections we list several properties of this model. We begin first by  a useful scaling relation of the  variance tensor. 
\subsection{Variance tensor scaling}\label{Scaling}
\label{Section-Scaling}
In the conditions of Kolmogorov-Richardson scaling, at scale $\ell$, the velocity increments and the eddies turn-over time scale as $u_\ell \sim \epsilon ^{1/3} \ell^{1/3}$ and $\tau_\ell \sim \epsilon ^{-1/3} \ell^{2/3}$ respectively, with $\epsilon$ denoting a constant energy dissipation rate across the inertial scales range. The turn-over time ratio for two different scales in this range,
$\frac{\tau_L}{\tau_\ell} \sim (\frac{L}{\ell})^{2/3}$, exhibits a direct relation between a change of time scale and a change of spacial resolution. A coarsening in time yields  thus a space dilation. 
  
Besides, at scale $L$, we have:
\begin{align}
\nonumber
\Exp (u_L-\Exp  u_L)^2\tau_L \sim \epsilon^{1/3} L^{4/3}.\nonumber
\end{align}
 At the smallest scale, this quantity corresponds  to the variance tensor $a_\ell= \Exp (u_\ell-\Exp u_\ell)^2 \tau_\ell$. 
 We have thus:
\begin{equation}
\label{eq1}
\frac{a_\ell}{ \Exp (u_L-\Exp u_L)^2\tau_L }= (\frac{\ell}{L})^{4/3},
\end{equation}
 Within a given cell, $V_L$, at scale $L$, the energy  of the small-scale random field  is given as:
\begin{equation}
\Exp \sum_{\x_i\in V(L)}\| \bsigma_\ell(\xx_i) \dif\B_t\|^2= a_L \tau_\ell=  (\frac{L}{\ell})^3 a_\ell \tau_\ell,
\end{equation}
which from (\ref{eq1}) gives us: 
\begin{align}
\nonumber
a_L&\sim (\frac{L}{\ell})^3 (\frac{\ell}{L})^{4/3} \Exp (u_L-\Exp u_L)^2\tau_L ,\\
\label{scaling}
&\sim  (\frac{L}{\ell})^{5/3} \Exp(u_L-\Exp u_L)^2\tau_L.
\end{align}
This relation  provides us  an expression of the scaling between the subgrid variance tensor at scale $L$ and the resolved velocity anomalies. It will serve us to impose a proper tuning of the subgrid term when the  variance tensor  is defined from the resolved velocity.   
\subsection{Adimensionalization of large-scale stochastic Navier-Stokes equations}
Considering the scaled coordinates $\xx^*= L \xx$, $t^*=Tt$, with velocity $\w^*=U\w$, pressure $p^*=U^2 p$ and a variance tensor  $\mbs a ^*=  A(L)\;\mbs a$, we have:
\begin{equation}
\partial_{t^*} \w^* + \w^*\nabla^* (\w^* -\frac{1}{2} \nab^*\bcdot \mbs a^*)= -\nabla^* p^* + \frac{1}{2} \sum_{i,j=1}^d \partial_{x^*_i}  (a_{ij}^*\partial_{x^*_j}  \w^*)+\nu \Delta^* \w^*,\nonumber
\end{equation}
\begin{multline}
\frac{U^2}{L}\partial_{t} \w + \frac{U^2}{L} \w\nabla\transp \w -\frac{1}{2} \frac{UA(L)}{L^2} \w\nabla\transp \nab\bcdot \mbs a =\nonumber\\ - \frac{U^2}{L}\nabla p + \frac{1}{2} \frac{UA(L)}{L^2} \sum_{i,j=1}^d \partial_{x_i}  (a_{ij}\partial_{x_j}  \w) + \frac{U}{L^2}\nu \Delta \w.\nonumber
\end{multline}
Assuming, as in the previous section, that the characteristic value of the resolved variance tensor, $A(L)$, depends linearly on the  variance tensor at the smallest scale,  $A(\ell)$, 
the dimensionless form of the large-scale stochastic Navier-Stokes equations  reads: 
\begin{equation}
\partial_{t} \w + \w\nabla\transp (\w - \frac{1}{\Upsilon} \nab\bcdot \mbs a) = - \nabla p + \frac{1}{\Upsilon} \sum_{i,j=1}^d \partial_{x_i} (a_{ij}\partial_{x_j}\w) + \frac{1}{R_e}\Delta \w,
\label{D-NSU}
\end{equation}
where we introduced a dimensionless number 
\begin{equation}
\Upsilon=  \frac{2 U^2}{K(L/\ell) \Sigma_{U}},
\end{equation}
which relates   the typical kinetic energy at scale $L$ to a characteristic value of the velocity variance, $\Sigma_U$, at the same level. In this expression we considered from (\ref{scaling}) a characteristic value of the resolved variance tensor
\begin{equation}
A(L)= K(\frac{L}{\ell}) \Sigma_U \frac{L}{U},
\end{equation} where we introduced a scale factor ratio, $K(\frac{L}{\ell})$ (with K(1)=1).  As previously inferred, under the Kolmogorov scaling hypothesis  this ratio scales as $(L/\ell)^{5/3}$. 
Let us finally observe that  for a typical  variance tensor, $A(\ell)$,  tending to zero, (\ref{D-NSU}) tends to the original Navier-Stokes equations. It is the situation occurring when  the  variance tensor is negligible in front of the drift energy.

In the next section we will see that the large-scale model proposed conserves the invariance properties of the original Navier-Stokes equations. 
\subsection{Invariance of large-scale stochastic Navier-Stokes equations}
Let us consider in the following the classical symmetry groups of the incompressible Navier-Stokes equations  in a periodic domain and in the absence of external forcing.
\subsubsection{Translation-invariance}
Space translation invariance is achieved if the whole velocity field ${\bf U}^*(\xx,t) = {\bf U}(\xx-{\bf b},t)$ is still a solution of the Navier-Stokes equations. In our setting we have
\[
\w^*(\xx,t)\dif t + \bsigma^*(\xx,t) \dif \B_t= {\w}(\xx-{\bf b},t)\dif t + \bsigma(\xx-{\bf b} ,t) \dif\B_t.
\]
Separating the Brownian component from the smooth term yields: ${\w}^*(\xx,t)= {\w}(\xx-{\bf b},t)$,  $\bsigma^*(\xx,t)= \bsigma(\xx-{\bf b},t)$ and hence $\partial_t {\w}^*= \partial_t {\w}$,  $\partial_{x_i}{\w}^*= \partial_{x_i} {\w}$, $\partial_{x_i}\bsigma^*= \partial_{x_i}\bsigma$, $\partial_{x_i} {\bf a}^*= \partial_{x_i}{\bf a}$, $\partial_{x_i}(a^*_{ij} \partial _{x_j} {\w}^*) = \partial_{x_i}(a_{ij} \partial_{x_j}{\w})$, and translational invariance follows immediately.
\subsubsection{Time-shift invariance}
Time-shift invariance is obtained recalling that Brownian motion has itself a time-shift  invariance property. This property allows us to write (in law) for $t^*= t-b$:
\[
\bsigma^*(\xx,t) \dif \B_{t^*} \inlaw \bsigma^*(\xx,t) \dif \B_{t} \defin\bsigma(\xx,t-b) \dif \B_{t}, 
\]
which leads straightforwardly to time-shift invariance.
%
%
%
%
%
\subsection{Rotational and reflectional invariance}
These two symmetry groups correspond to a constant rotation or to a reflection of the coordinates system. The transformed coordinates are $x_i^*= \sum_j R_{ij} x_j$ and $U_i^*= \sum_j R_{ji} U_j(\xx^*)$. 
Note that $\bsigma^* \dif\B_t= {\mbs R}\transp\bsigma(\xx^*) \dif\B_t$, which yields ${\bf a}^*(\xx)= {\mbs R}\transp \bsigma(\xx^*) \bsigma(\xx^*)\transp {\mbs R}= {\mbs R}\transp {\bf a} (\xx^*){\mbs R}$ and hence $ \sum_{i,j=1}^d \partial_{x_i} (a^*_{ij} \partial_{x_j}\w^*)=  {\mbs R}\transp \sum_{i,j=1}^d \partial_{x_i}(a_{ij} (\xx^*)\partial_{x_j}\w(\xx^*)) $. The invariance of the transformed Navier-Stokes equations follows immediately:
\begin{multline}
\mbs R \transp \partial_t \w (\xx^*) + \mbs R \transp \w(\xx^*) \nab\transp \bigl( \w(\xx^*)-\frac{1}{2}\nab\bcdot \mbs a(\xx^*)\bigr)  =\\ \mbs R\transp \sum_{i,j=1}^d \partial_{x_i} \bigl(a_{ij} (\xx^*)_{x_j}\w(\xx^*)\bigr)  - \mbs R \transp\nab p(\xx^*) +\nu \mbs R\transp \Delta \w(\xx^*).
\end{multline}
\subsubsection{Galilean transformations invariance}
A model is said to be Galilean invariant if it is invariant with respect to an inertial transformation of the representation frames: $\xx^*= \xx- {\bf V} t,\; t^*=t$. The velocity in this translated non-accelerating system of reference reads 
\[
\w^*\dif t + \bsigma^*(\xx,t)\dif\B_{t}= \w(\xx-{\bf V}t,t) \dif t+ {\bf V}\dif t + \bsigma(\xx-{\bf V}t,t)\dif \B_{t},
\]
with 
\[
{\mbs a}^* (\xx,t) = \bsigma(\xx^*,t) \bsigma(\xx^*,t) = {\mbs a} (\xx^*,t).
\]
From those equations we obtain: 
\begin{align}
\partial_{x_j} w_i^* &= \partial_{x_j} w_i,\\
\partial_{t} w^* &= \partial_{t} w - \partial_{x_j} w^i V^j, \\
\partial_{x_i} (a_{ij}^*\partial_{x_j} \w^*) &=  \partial_{x_i} (a_{ij}\partial{x_j}{\w}).
\end{align}
As the pressure and  viscous force are Galilean invariant and since $\frac{\dif}{\dif t} \w^* = \frac{\dif}{\dif t} \w$, this shows the system  is  Galillean invariant. 
\subsubsection{Time reversal invariance}
As  with the original Navier-Stokes equation, the stochastic model  is not invariant to a time reversal transformation: $\w^*(\xx, t) = -\w (\xx,-t)$. This property is respected only for the Euler equation. Despite the time reversibility property of Brownian motion this loss of symmetry is due to  the even nature of the quadratic tensor.
\subsubsection{Scale invariance}
Considering the scaled transformation: $$\w^*(\xx,t)\dif t  + \bsigma^*(\xx,t)\dif \B_{t^*}= \lambda^h \w(\lambda^{-1}\xx,\lambda^{h-1}t) \dif t+ \lambda^h\boldsymbol{\sigma} (\lambda^{-1}\xx,\lambda^{h-1}t) \dif\B_{t^*},$$ we get:
\begin{align}
&\partial_{t} {\w}^*= \lambda^{2h-1} \partial_{t}{\w},\\
&\partial_{x_j} w^*_i w^*_j = \lambda^{2h-1} \partial_{x_j} w_i w_j,\\
&\nab p^* =  \lambda^{2h-1} \nab p,\\
&\bsigma ^*(\xx,t)\dif\B_{t^*} =\lambda ^{\frac{1}{2}(1-h)}\bsigma^*(\xx,t)\dif \B_t= \lambda ^{\frac{1}{2}(h+1)} \bsigma (\xx^*,t^*)\dif \B_t,\\
& {\bf a}^*=  \lambda ^{h+1} {\bf a} (\lambda^{-1}\xx,\lambda^{h-1}t),\\
&\partial_{x_j} w^*_i \partial_{x_k}\a_{kj}^* = \lambda^{2h-1} \partial_{x_j} w_i\partial_{x_k}\a_{kj}\\
&\partial_{x_i} (a^*_{ij} \partial_{x_j}{\w}^*) = \lambda^{2h-1} \partial_{x_i} (a_{ij} \partial_{x_j}{\w}),\\
&\Delta {\w}^* = \lambda^{h -2}\Delta \w.
\end{align}
The two last relations highlight a very interesting property of the modified Navier-Stokes equations. When the viscosity can be neglected in comparison to the variance tensor, the large-scale stochastic Navier-Stokes equations  become scale-invariant for any values of the scale exponent. They share this property with the Euler equations.  When the friction term is considered the equations are invariant only for $h=-1$.

To sum up,  the large-scale stochastic representation of the Navier-Stokes equations  have the remarkable feature of keeping all the symmetries of the original equations and to follow the same scale invariance property as the Euler equation when neglecting the friction term. This last property,  which inherits from the time scale invariance of Brownian motion, is often not verified for large-scale eddy viscosity representations of the Euler equations. 


%
%
%
%
%

\section{Gradient of the functional ${\cal J}$}\label{GradientComputation}
Using the condition $\nab\cdot(\nab\cdot\a)=0$, we can write:
\begin{equation*}
{\cal J(\a)}=\frac{1}{2}\| \bar\w_{ {\boldsymbol \ell}}{\nab}\transp \w_{ {\boldsymbol \ell}}+\w_{ {\boldsymbol \ell'}}{\nab}\transp \bar\w_{ {\boldsymbol \ell}}-\frac{1}{2}\sum_{i,j=1}^d\partial_{x_i} \partial_{x_j}\bigl(a_{ij}(\w_{ {\boldsymbol \ell}}-\lambda \w_{ {\boldsymbol \ell'}})\bigr)\|^2_{L^2(\R^d)},
\end{equation*}
where $\bar\w_{ {\boldsymbol \ell}}=\w_{ {\boldsymbol \ell}}-\w_{ {\boldsymbol \ell'}}$ is the multi-scale residual term. Then, it is easy to see that:
\begin{multline*}
\frac{1}{\epsilon}[{\cal J}(\a+\epsilon a_{mn}\Id )-{\cal J}(\a)]\xrightarrow[\epsilon\to 0]{}
\\
-\frac{1}{2}\langle \partial_{x_m} \partial_{x_n}[a_{mn}(\w_{ {\boldsymbol \ell}}-\lambda \w_{ {\boldsymbol \ell'}})], \bar\w_{ {\boldsymbol \ell}}{\nab}\transp \w_{ {\boldsymbol \ell}}+\w_{ {\boldsymbol \ell'}}{\nab}\transp \bar\w_{ {\boldsymbol \ell}}-\frac{1}{2}\sum_{i,j=1}^d\partial_{x_i} \partial_{x_j}[a_{ij}(\w_{ {\boldsymbol \ell}}-\lambda \w_{ {\boldsymbol \ell'}})]\rangle,
\end{multline*}
and integration by part gives:
\begin{multline*}
\partial_{a_{mn}}{\cal J}(\a)= \\-\frac{1}{2}(\w_{ {\boldsymbol \ell}}-\lambda \w_{ {\boldsymbol \ell'}})\partial_{x_m} \partial_{x_n}\biggl(\bar\w_{ {\boldsymbol \ell}}{\nab}\transp \w_{ {\boldsymbol \ell}}+\w_{ {\boldsymbol \ell'}}{\nab}\transp \bar\w_{ {\boldsymbol \ell}}-\frac{1}{2}\sum_{i,j=1}^d\partial_{x_i} \partial_{x_j}[a_{ij}\bigl(\w_{ {\boldsymbol \ell}}-\lambda \w_{ {\boldsymbol \ell'}})]\biggr).
\end{multline*}
Similarly, the derivative $\partial_{\lambda}{\cal J}$ with respect to the constant $\lambda$ is given by: 

\begin{multline*}
\partial_{\lambda}{\cal J}(\a)= \\\frac{1}{2}\sum_{m,n=1}^da_{mn}\w_{ {\boldsymbol \ell'}}\cdot\partial_{x_m} \partial_{x_n}\biggl(\bar\w_{ {\boldsymbol \ell}}{\nab}\transp \w_{ {\boldsymbol \ell}}+\w_{ {\boldsymbol \ell'}}{\nab}\transp \bar\w_{ {\boldsymbol \ell}}-\frac{1}{2}\sum_{i,j=1}^d\partial_{x_i} \partial_{x_j}[a_{ij}\bigl(\w_{ {\boldsymbol \ell}}-\lambda \w_{ {\boldsymbol \ell'}})]\biggr).
\end{multline*}

\end{document}